\documentclass[12pt,english,floatfix,superscriptaddress,aps,prd,preprint]{revtex4}
\usepackage[utf8]{inputenc}
\usepackage{amsmath}
\usepackage{amssymb}
\usepackage{amsbsy}
\usepackage{amsfonts}
\usepackage{amsopn}
\usepackage{amstext}
\usepackage{graphicx}
\usepackage{amssymb}
\usepackage{amsfonts}
\usepackage{amsmath}
\usepackage{braket}
\usepackage{amsmath,amsthm,amsfonts,amssymb}
\usepackage[mathcal]{eucal}
\usepackage{mathrsfs}
\usepackage{graphicx}
\usepackage[english]{babel}
\usepackage{color}
\usepackage{esint}
\usepackage[dvips]{epsfig}
\usepackage[dvips]{graphicx}
\usepackage{float}
\usepackage{units}
\usepackage{textcomp}

\usepackage{hyperref}
\usepackage{slashed}

\newcommand{\ie}{\begin{equation}}
\newcommand{\fe}{\end{equation}}
\newcommand{\se}{\begin{eqnarray}}
\newcommand{\ff}{\end{eqnarray}}

\begin{document}

\title{The noncommutative Dirac oscillator with a permanent electric dipole moment in the presence of an electromagnetic field}


\author{R. R. S. Oliveira}
\email{rubensrso@fisica.ufc.br}
\affiliation{Universidade Federal do Cear\'a (UFC), Departamento de F\'isica,\\ Campus do Pici, Fortaleza - CE, C.P. 6030, 60455-760 - Brazil.}
\author{G. Alencar}
\email{geova@fisica.ufc.br}
\affiliation{Universidade Federal do Cear\'a (UFC), Departamento de F\'isica,\\ Campus do Pici, Fortaleza - CE, C.P. 6030, 60455-760 - Brazil.}
\author{R. R. Landim}
\email{renan@fisica.ufc.br}
\affiliation{Universidade Federal do Cear\'a (UFC), Departamento de F\'isica,\\ Campus do Pici, Fortaleza - CE, C.P. 6030, 60455-760 - Brazil.}
\date{\today}

\begin{abstract}

In this paper, we investigate the bound-state solutions of the noncommutative Dirac oscillator with a permanent electric dipole moment in the presence of an electromagnetic field in (2+1)-dimensions. We consider a radial magnetic field generated by anti-Helmholtz coils, and the uniform electric field of the Stark effect. Next, we determine the bound-state solutions of the system, given by the two-component Dirac spinor and the relativistic energy spectrum. We note that this spinor is written in terms of the generalized Laguerre polynomials, and this spectrum is a linear function on the potential energy $U$, and depends explicitly on the quantum numbers $n$ and $m$, spin parameter $s$, and of four angular frequencies: $\omega$, $\tilde{\omega}$, $\omega_\theta$, and $\omega_\eta$, where $\omega$ is the frequency of the oscillator, $\tilde{\omega}$ is a type of ``cyclotron frequency'', and $\omega_\theta$ and $\omega_\eta$ are the noncommutative frequencies of position and momentum. Besides, we discussed some interesting features of such a spectrum, for example, its degeneracy, and then we graphically analyze the behavior of the spectrum as a function of the four frequencies for three different values of $n$, with and without the influence of $U$. Finally, we also analyze in detail the nonrelativistic limit of our results, and comparing our problem with other works, where we verified that our results generalize several particular cases of the literature.

\end{abstract}

\maketitle


\section{Introduction}

In 1989, M. Moshinsky and A. Szczepaniak developed the first relativistic version of the quantum harmonic oscillator (QHO) for spin-1/2 fermions, in which it became known as the Dirac oscillator (DO) \cite{Moshinsky,Strange}. To build the DO (an exactly soluble model), we must to insert into the free Dirac equation (DE) a nonminimal coupling given by: ${\bf p}\rightarrow {\bf p}-im_{0}\omega\beta {\bf r}$, where ${\bf p}=-i\hbar\boldsymbol{\nabla}$ is the momentum operator, $m_{0}>0$ is the rest mass, $\omega>0$ is the angular frequency, $\beta$ is one of the usual Dirac matrices, and {\bf{r}} is the position vector (operator) \cite{Moshinsky,Martinez,Strange}. In particular, such an interaction term (or coupling term) it is not interesting just because it generates the QHO with a strong spin-orbit coupling in the nonrelativistic limit, but because it may also have applications in quantum chromodynamics (QCD) \cite{Bentez,Martinez,Moreno}. For example, since the interaction or confinement potential (the well-known Cornell potential) has a term linear in the position, implies that the interaction term of the DO could serve to model quark confinement in QCD \cite{Bentez,Martinez,Moreno,Lucha}. In addition, it has already been shown that such an interaction term arises through the DE to a nucleon (the neutron, for example) with a (anomalous) magnetic dipole moment (MDM) interacting with an external electromagnetic field (actually, a radial electric field, or an effective chromoelectric field) \cite{Bentez,Martinez,Moreno,BJP}.

Since it was introduced in the literature, several publications on the DO have been carried out in different areas of physics, such as in thermodynamic \cite{Pacheco1,Pacheco2,Boumali1}, physics-mathematics \cite{Bentez,Chargui,Rao,Szmytkowski}, quantum optics \cite{Bermudez1,Bermudez2,Rozmej,Longhi}, graphene physics (or condensed matter physics) \cite{Quimbay,Boumali1,Belouad,Neto,Cunha}, etc, and has also been studied in different contexts, such as in problems involving quantum phase transitions \cite{Bermudez3,Menculini}, hidden supersymmetry \cite{Bentez}, Aharonov-Bohm-Coulomb system \cite{Oliveira1}, spin effects \cite{Andrade1,Andrade2,Villalba}, rotating frames in the cosmic string spacetime \cite{Oliveira2,Oliveira3}, minimal length scenario \cite{Menculini}, two-dimensional relativistic quantum rings \cite{Neto,Bakke1}, etc. In 2013, the DO in (1+1)-dimensions was verified experimentally by J. A. Franco-Villafa\~ne et al \cite{Franco}, where the experimental apparatus was based on a microwave system consisting of a chain of coupled dielectric disks. However, in addition to DO, there are also other types of relativistic oscillators in the literature (but not yet experimentally proven), such as the Klein-Gordon oscillator (for spin-0 bosons) \cite{Bruce} and the Duffin-Kemmer-Petiau oscillator (for spin-0 and spin-1 bosons) \cite{Nedjadi}, respectively. Recently, the DO has been studied in connection with Lorentz symmetry violation \cite{Vitoria}, Aharonov–Bohm and Aharonov–Casher effects \cite{Ahmed,Candemir}, thermal properties \cite{Korichi}, and with position-dependent mass \cite{Schulze}.

In 1947, H. S. Snyder published some papers on quantized spacetimes, thus introducing the concept of noncommutative space (NCS) or noncommutative spacetime (NCST) as we know it today \cite{Snyder1,Snyder2}. For Snider, although the Minkowski spacetime is a continuum, this assumption is not required by Lorentz Invariance. In this way, Snider proposed a model of a Lorentz invariant discrete spacetime inspired by quantum mechanics. In Refs. \cite{Namsrai1,Sidharth}, some interesting consequences of a quantized spacetime are discussed, particularly, for a grand unified theory. Already in Refs. \cite{Szabo,Majid,Abel,Pikovski,Moffat}, a NCS is considered a possible scenario for the short-distance behavior (Planck length scale) of some physical theories (quantum gravity, for example). So, in order to generalize the already well-established NCS, the concept of noncommutative phase space (NCPS) was later introduced, where now both obey a rigorous mathematical formulation: the noncommutative geometry (NCG) \cite{Szabo,Douglas,Gomis,Seiberg}. In essence, a NCPS is based on the assumption that the position and momentum operators are NC variables and now must satisfy $[x^{NC}_\mu,x^{NC}_\nu]\neq{0}$ and $[p^{NC}_\mu,p^{NC}_\nu]\neq{0}$. For more details on the phenomenology of NCG, please see Refs. \cite{Hinchliffe,Melic,Schupp,Abel,Pikovski}, where supposed signatures of NC were investigated in decay of kaons, photon-neutrino interaction, etc. Also, a NCS (NCSP) has a wide range of applications, such as in QCD \cite{Carlson}, quantum electrodynamics (QED) \cite{Riad}, black holes \cite{Nicolini}, quantum cosmology \cite{Garcia}, Shannon entropy \cite{Nascimento}, graphene \cite{Bastos}, quantum wells \cite{Berto}, and in the DO and QHO \cite{Boumali2,Giri}, etc, and recently was applied in  the relativistic and nonrelativistic quantum Hall effect \cite{Oliveira4}, and of the Pauli oscillator (PO) \cite{Heddar}.

In classical electrodynamics, an electric dipole moment (EDM) is defined from the first-order term of the multipole expansion of the electrostatic potential for a discrete or continuous charge distribution; and consists of two equal and opposite charges that are infinitesimally close \cite{Jackson,Lacava}. Mathematically, such an EDM is defined as ${\bf d}\equiv q{\bf r}$ (discrete case), where $q$ is the total charge, or ${\bf d}=\int {\bf r}'\rho({\bf r}')dV'$ (continuous case), where $\rho({\bf r}')$ is the charge density \cite{Jackson,Lacava}. However, in some atoms (Xe-129, Hg-199, Ra-225, $\ldots$) \cite{Vold,Allmendinger,Romalis,Parker,Dzuba} and molecules (H2O, HCl, CO, $\ldots$) \cite{Jackson,Lacava} in which the centers of positive and negative charge do not coincide, then there will be a permanent EDM, given by ${\bf d}_0=d_0\hat{\bf n}$, where $\hat{\bf n}=\frac{\bf n}{\vert \bf n \vert}$ is a unit vector (can be oriented in any direction in space). In particular, such an EDM (intrinsic or non-null EDM) is already ``born'' with the atom or molecule itself. Also, even in atoms and molecules with a null EDM, the action of an external electric field can separate the centers of positive and negative charges and thus produce an induced EDM \cite{Jackson,Lacava}, where is defined as ${\bf d}\equiv\alpha{\bf E}$, where $\alpha$ is the polarizability and ${\bf E}$ is the electric field \cite{Jackson,Lacava}.

Now, from the point of view of elementary particle physics, nuclei and particles (Dirac fermions) ``can'' also have a non-null permanent EDM, given by ${\bf d}_f$ and defined as: ${\bf d}_f\equiv d_f\hat{\bf N}$ ($d_f=const.$), where $\hat{\bf N}=\frac{\bf J}{\vert \bf J \vert}$, being ${\bf J}$ the spin angular momentum, i.e., ${\bf d}_f$ must be parallel or antiparallel to the angular momentum (cannot be oriented in any direction in space) \cite{Engel,Chupp,Commins}. In particular, here an EDM cannot exist unless both parity (P) and time-reversal (T) invariance are violated, where both nonrelativistic and relativistic Hamiltonians for the EDM, given by $H_{EDM}^{Pauli}=-{\bf d}_f \cdot {\bf E}=-d_f\frac{\bf J}{\vert \bf J \vert}\cdot {\bf E}$ (Pauli Hamiltonian) and $H_{EDM}^{Dirac}=-d_f\gamma^0{\bf \Sigma}\cdot {\bf E}$ (Dirac Hamiltonian), are odd under P and T transformations \cite{Engel,Chupp,Commins}. However, due to current experimental upper limits, where EDMs must be extremely small, implies that so far no EDMs have been observed (unlike classical electrodynamics) \cite{Engel,Chupp,Commins,Pospelov,Bolokhov}. But as has already been said, EDMs may be non-null because P and T are in fact violated in nature \cite{Engel,Chupp,Commins,Pospelov,Bolokhov}. For example, if we assume CPT invariance as an exact symmetry, then the T-violation is equivalent to CP violation (observed in meson decays), i.e., EDM is a direct signal of T-violation and also of CP violation. \cite{Engel,Chupp,Commins,Pospelov,Bolokhov}. Thus, CP violation could act to generate EDMs via radiative corrections \cite{Engel,Chupp,Commins,Pospelov,Bolokhov}. Also, EDMs have already been employed for a possible search for dark matter \cite{Heo}, and for a possible explanation of the baryonic asymmetry problem (matter-antimatter asymmetry in the universe), since the Standard Model cannot produce the observed asymmetry (new CP violatings are required) \cite{Pospelov,Bolokhov}.

The present paper has as its goal to investigate the bound-state solutions of the NCDO with a permanent EDM in the presence of an external electromagnetic field in the (2+1)-dimensional Minkowski spacetime (a problem not yet studied in the literature, so far). For simplicity and without loss of generality, in this work we will just call it EDM. With respect to the electromagnetic field, here we consider a radial magnetic field and linear at $r$ (or $x$ and $y$) generated by anti-Helmholtz coils, and the uniform electric field of the Stark effect, in which it is described by a uniform electric field in the z-direction. Now, with respect to bound-state solutions, we mainly focus on the solutions of an eigenvalue equation, given by the eigenfunctions (Dirac spinor and spinorial wave function) and on energy eigenvalues (relativistic and nonrelativistic energy spectrum). In this work, we also consider the ``spin'' of the planar fermion, described by a parameter $s$, called spin parameter (is not the spin quantum number), where $s=+1$ is for the spin ``up'', and $s=-1$ is for the spin ``down''. In this way, we can say that we are going to investigate the bound-state solutions under the influence of spin effects, NC effects, and electromagnetic effects, respectively.

The structure of this work is organized as follows. In Sect. \ref{sec2}, we briefly review the formalism on the NCPS in two-dimensions (nonrelativistic case), and in $(2+1)$-dimensions (relativistic case). In Sect. \ref{sec3}, we start our discussion by initially introducing the total Lagrangian (density) of the system, which is the sum of two Lagrangians: the free Dirac Lagrangian, and the Lagrangian for neutral fermions with EDM. From this total Lagrangian, we obtain the equation of motion of the NCDO with EDM, both in linear and quadratic form. In Sect. \ref{sec4}, we analyze the asymptotic behavior of our second-order differential equation (radial or quadratic NCDO), where we get the bound-state solutions, given by the NC Dirac spinor and the relativistic energy spectrum. In Sect. \ref{sec5}, we analyze the nonrelativistic limit of our results, where we get the equation of motion for the nonrelativistic NCDO as well as the nonrelativistic bound-state solutions, given by the spinorial wave function and the nonrelativistic spectrum. Finally, in Sect. \ref{sec6} we present our conclusions and some remarks.

\section{The two-dimensional noncommutative phase space}\label{sec2}

In usual two-dimensional (2D) quantum mechanics, a quantum phase space (commutative phase space) is defined by substituting the classical canonical variables of position and momentum ($x_i$ and $p_j$) by their respective Hermitian (quantum) operators, now written as $\hat{x}_i$ and $\hat{p}_j$, where obey the following Heisenberg (canonical) commutation relations (usual or ordinary Heisenberg algebra) \cite{Szabo}
\begin{equation}\label{NC1}
[\hat{x}_i,\hat{p}_j]=i\hbar\delta_{ij}, \ \ [\hat{x}_i,\hat{x}_j]=0,\ \
[\hat{p}_i,\hat{p}_j]=0, \ \ (i,j=1,2=x,y),
\end{equation}
and whose Heisenberg uncertainty relations (Heisenberg uncertainty principle for position and momentum) are written as follows
\begin{equation}\label{NC2}
\Delta\hat{x}_i\Delta\hat{p}_j\geq\frac{\hbar}{2}\delta_{ij}, \ \ \Delta\hat{x}_i\Delta\hat{x}_j=0, \ \ \Delta\hat{p}_i\Delta\hat{p}_j=0,
\end{equation}
where $\delta_{ij}=\delta_{ji}$ is the Kronecker delta (Euclidean metric). In essence, the uncertainty principle states that we cannot measure simultaneously and with high precision two operators that do not commute with each other (incompatible operators), and therefore, the more we know about one operator (or observable), the less we know about the other (and vice versa).

Now, in order to define a noncommutative quantum phase space (NCQPS), or simply a NCPS \cite{Berto,Bastos,Nascimento}, the relations given in \eqref{NC1} must then obey the following deformed Heisenberg (noncanonical) commutation relations (NC or deformed Heisenberg algebra)
\begin{equation}\label{NC3}
[\hat{x}^{NC}_i,\hat{p}^{NC}_j]=i\hbar\delta_{ij}\left(1+\frac{\theta\eta}{4\hbar^2}\right), \ \ [\hat{x}^{NC}_i,\hat{x}^{NC}_j]=i\theta_{ij},\ \
[\hat{p}^{NC}_i,\hat{p}^{NC}_j]=i\eta_{ij},
\end{equation}
where the NC operators $\hat{x}^{NC}_i$ and $\hat{p}^{NC}_i$ are defined as
\begin{equation}\label{NC4}
\hat{x}^{NC}_i\equiv\hat{x}_i-\frac{1}{2\hbar}\theta_{ij}\hat{p}_j, \ \ \hat{p}^{NC}_i\equiv\hat{p}_i+\frac{1}{2\hbar}\eta_{ij}\hat{x}_j, \ \ (\hat{x}_i=\delta_{ij}\hat{x}^j; \ \hat{p}_j=\delta_{ij}\hat{p}^i=-i\partial_j),
\end{equation}
with $\theta_{ij}\equiv\theta\epsilon_{ij}$ and $\eta_{ij}\equiv\eta\epsilon_{ij}$ being antisymmetric constants ``tensors'' (real deformation parameters), $\epsilon_{ij}=-\epsilon_{ji}$ is the Levi-Civita symbol (a pseudotensor), and $\theta>0$ and $\eta>0$ ($\theta,\eta\in\mathbb{R}^\star_+$) are the position and momentum NC parameters (NC parameter for the coordinate and momentum space) with dimensions of length-squared and momentum-squared, respectively. From a phenomenological point of view, these two NC parameters can have the following values: $\theta\leq 4.0 \times 10^{-40}$m$^2$ and $\eta\leq 1.76 \times 10^{-61}$kg$^2$m$^2$s$^{-2}$ \cite{Berto}.

In particular, the relationship between the set of NCs variables $\{\hat{x}^{NC}_i, \hat{p}^{NC}_i\}$ (here is not the anticommutator) with the set of usual variables $\{\hat{x}_i, \hat{p}_i\}$ is a consequence of a noncanonical linear transformation, known as Darboux transformation or Seiberg-Witten map \cite{Berto,Bastos,Nascimento}. However, all physical observables are entirely independent of the chosen map (how it should be). Furthermore, the NCS (NC only in position) causes a change in the usual product of two arbitrary functions $F({\bf r})$ and $G({\bf r})$, where now such a product is called of star product or Moyal product, and whose definition is given as follows \cite{Berto}
\begin{equation}\label{NC5}
F({\bf r},\theta)\star G({\bf r},\theta)\equiv F({\bf r})e^{(i/2)(\overleftarrow{\partial}_{x_i}\theta_{ij}\overrightarrow{\partial}_{x_j})}G({\bf r})=F({\bf r})e^{(i\theta/2)(\overleftarrow{\partial}_{x}\overrightarrow{\partial}_{y}-\overleftarrow{\partial}_{y}\overrightarrow{\partial}_{x})}G({\bf r}).
\end{equation}

In fact, in the absence of the NCS ($\theta_{ij}=0$), the star product is simply the usual product $F({\bf r})G({\bf r})$. Also, the uncertainty relations for the NC case are now written as
\begin{equation}\label{NC6}
\Delta\hat{x}^{NC}_i\Delta\hat{p}^{NC}_j\geq\frac{\hbar_{eff}}{2}\delta_{ij}, \ \ \Delta\hat{x}^{NC}_i\Delta\hat{x}^{NC}_j\geq\frac{1}{2}\vert\theta_{ij}\vert, \ \ \Delta\hat{p}^{NC}_i\Delta\hat{p}^{NC}_j\geq\frac{1}{2}\vert\eta_{ij}\vert,
\end{equation} 
where
\begin{equation}\label{NC7}
\hbar_{eff}=\hbar^{NC}=\hbar(1+\xi),
\end{equation}
with $\hbar_{eff}$ being the effective, deformed or NC Planck constant, and the parameter $\xi$ is given by: $\xi\equiv\frac{\theta\eta}{4\hbar^2}$. For $\xi\ll 1$, or in the (commutative) limit $\xi\to 0$, we recover the usual uncertainty relations. For a more detailed discussion of the possible hypothetical values of $\xi$, please see Ref. \cite{Berto}, where the NC gravitational quantum well was studied.

Last but not least, the NCPS can also be expanded to include the $(2+1)$-dimensional Minkowski spacetime of the relativistic quantum mechanics \cite{Berto}, which is given by following relations
\begin{equation}\label{NC8}
[\hat{x}^{NC}_\mu,\hat{x}^{NC}_\nu]=i\hbar\left(\delta_{\mu\nu}+\frac{1}{4\hbar^2}\theta_{\mu}^{\sigma}\eta_{\nu\sigma}\right), \ \ [\hat{x}^{NC}_\mu,\hat{x}^{NC}_\nu]=i\theta_{\mu\nu},\ \
[\hat{p}^{NC}_\mu,\hat{p}^{NC}_\nu]=i\eta_{\mu\nu},
\end{equation}
where
\begin{equation}\label{NC9}
\hat{x}^{NC}_\mu=\hat{x}_\mu-\frac{1}{2\hbar}\theta_{\mu\nu}\hat{p}^\nu, \ \ \hat{p}^{NC}_\mu=\hat{p}_\mu+\frac{1}{2\hbar}\eta_{\mu\nu}\hat{x}^\nu, \ \ (\hat{x}_\mu=g_{\mu\nu}\hat{x}^\nu; \ \hat{p}_\mu=g_{\mu\nu}\hat{p}^\nu),
\end{equation}
and
\begin{equation}\label{NC10}
F({\bf r},\theta)\star G({\bf r},\theta)\equiv F({\bf r})e^{(i/2)(\overleftarrow{\partial}_{x^{\mu}}\theta^{\mu\nu}\overrightarrow{\partial}_{x^{\nu}})}G({\bf r}),
\end{equation}
being $g_{\mu\nu}=g^{\mu\nu}$= diag$(1,-1,-1)$ the flat metric tensor (Minkowski metric) and $\mu,\nu=0,1,2=t,x,y$. In addition, in this work we consider the NC only in the spatial components of $\hat{x}^{NC}_{\mu}$ and $\hat{p}^{NC}_{\nu}$ (space-like NC or spatial NC), where it implies: $\theta_{0i}=\eta_{0j}=0$; otherwise, the unitary, causality or locality of the quantum mechanics would not be preserved \cite{Berto,Szabo,Gomis}.

\section{The noncommutative Dirac oscillator in the (2+1)-dimensional Minkowski spacetime}\label{sec3}

To obtain the equation of motion of the NCDO with EDM in the presence of an external electromagnetic field (i.e., linear and quadratic NCDO), we will first derive the DE from the total Lagrangian (density) of our problem. Therefore, in quantum field theory (QFT) the Lagrangian for the free Dirac field (spin-1/2 free relativistic fermions) in (3+1)-dimensions is written as follows (Cartesian coordinates in SI units) \cite{Greiner}
\begin{equation}\label{1}
\mathcal{L}_0=\bar{\Psi}(i\hbar c \gamma^\mu \partial_\mu-m_0 c^2)\Psi, \ \ (\mu=0,1,2,3),
\end{equation}
where $\Psi=\Psi_D(t,{\bf r})=\ket{\Psi}=(\Psi_1,\Psi_2,\Psi_3,\Psi_4)^T \in\mathbb{C}^4$ is the four-component Dirac spinor (four-element column vector or Dirac field), and it is interpreted as a superposition of a spin-up/down particle, and a spin-up/down antiparticle, $\bar{\Psi}\equiv\Psi^{\dagger}\gamma^0$ is the adjunct spinor (Dirac adjunct) of $\Psi$, and $\gamma^\mu=(\gamma^0,\boldsymbol{\gamma})\equiv(\beta,\beta\boldsymbol{\alpha})$ are the gamma matrices ($\beta$ and $\boldsymbol{\alpha}$ are the usual Dirac matrices), which satisfies the anticommutation relation of the Clifford Algebra: $\{\gamma^\mu,\gamma^\nu\}=2g^{\mu\nu}I$, being $I$ a 4$\times$4 unit matrix. Explicitly, $\gamma^0$ (Hermitian), $\boldsymbol{\gamma}$ (anti-Hermitian), and $\boldsymbol{\alpha}$ (Hermitian) are defined in the standard Dirac representation as
\begin{equation}\label{Diracmatrices}
\gamma^0=\left(
    \begin{array}{cc}
      1\ &  \ 0 \\
      0\ &  -1 \\
    \end{array}
  \right)=\sigma_3 \otimes 1, \ \  \boldsymbol{\gamma}=\left(
    \begin{array}{cc}
      \ 0 & \ \boldsymbol{\sigma}  \\
      -\boldsymbol{\sigma} & \ 0 \\
    \end{array}
  \right)=i\sigma_2 \otimes \boldsymbol{\sigma}, \ \  \boldsymbol{\alpha}=\left(
    \begin{array}{cc}
      0 & \ \boldsymbol{\sigma} \\
      \boldsymbol{\sigma} & \ 0 \\
    \end{array}
  \right)=\sigma_1 \otimes \boldsymbol{\sigma},
\end{equation}
where $\boldsymbol{\sigma}=(\sigma_1,\sigma_2,\sigma_3)$ are the 2$\times$2 Pauli matrices, in which satisfies: $\{\sigma_i,\sigma_j\}=2\delta_{ij}$ and $[\sigma_i,\sigma_j]=2i\epsilon_{ijk}\sigma_k$, with $\sigma_i \sigma_j=\delta_{ij}+i\epsilon_{ijk}\sigma_k$, the 1’s and 0’s are 2$\times$2 unit and zero matrices, and $\otimes$ is the Kronecker product\cite{Sakurai}. Explicitly, the Pauli matrices take the form
\begin{equation}\label{Pauli}
\sigma_1=\sigma_x=\left(
    \begin{array}{cc}
      0\ &  1 \\
      1\ & 0 \\
    \end{array}
  \right), \ \  \sigma_2=\sigma_y=\left(
    \begin{array}{cc}
      0 & -i  \\
      i & \ 0 \\
    \end{array}
  \right), \ \  \sigma_3=\sigma_z=\left(
    \begin{array}{cc}
      1 & \ 0 \\
      0 & -1 \\
    \end{array}
  \right).
\end{equation}

Now, considering neutron fermions (neutrons, for example) with an permanent EDM given by $d_f=\pm \vert d_f \vert$, the Lagrangian is written as follows \cite{Engel,Chupp,Commins}
\begin{equation}\label{2}
\mathcal{L}_{EDM}=-i\frac{d_f}{2}\bar{\Psi}\sigma^{\mu\nu}\gamma_5 F_{\mu\nu}\Psi, \ \ (\mu,\nu=0,1,2,3),
\end{equation}
where $\sigma^{\mu\nu}=\frac{i}{2}[\gamma^\mu,\gamma^\nu]=i\gamma^\mu \gamma^\nu$ is an antisymmetric tensor (Hermitian tensor), $F_{\mu\nu}=\partial_\mu A_\nu-\partial_\nu A_\mu$ is the electromagnetic field tensor (electromagnetic field strength) with $A_\mu=(A_0/c,-{\bf A})$ being the electromagnetic four-potential (potential four-vector), and $\gamma_5=\gamma^5\equiv i\gamma^0\gamma^1\gamma^2\gamma^3$ is the fifth gamma matrix (Hermitian matrix), also called of chirality or handedness matrix, and satisfies: $\{\gamma^5,\gamma^\mu\}=0$, and $(\gamma^5)^2=I$. Explicitly, this matrix is defined as follows (in Dirac representation)
\begin{equation}\label{fifthtrices}
\gamma_5=\left(
    \begin{array}{cc}
      0 &  1 \\
      1 &  0 \\
    \end{array}
  \right)=\sigma_1\otimes 1.
\end{equation}

Therefore, the total Lagrangian is written as
\begin{equation}\label{3}
\mathcal{L}_T=\mathcal{L}_0+\mathcal{L}_{EDM}=\bar{\Psi}\left(i\hbar c \gamma^\mu \partial_\mu-i\frac{d_f}{2}\sigma^{\mu\nu}\gamma_5 F_{\mu\nu}-m_0 c^2\right)\Psi.
\end{equation}

Using now the Euler-Lagrange equation for $\bar{\Psi}$, given by \cite{Greiner}
\begin{equation}\label{4}
\partial_\mu\frac{\partial\mathcal{L}_T}{\partial(\partial_\mu\bar{\Psi})}-\frac{\partial\mathcal{L}_T}{\partial\bar{\Psi}}=0,
\end{equation}
we obtain as a result the following tensorial DE for a neutral fermion with EDM
\begin{equation}\label{5}
\left(\gamma^\mu p_\mu-i\frac{d_{f}}{2c}\sigma^{\mu\nu}\gamma_5 F_{\mu\nu}-m_0 c\right)\Psi=0,
\end{equation}
where $p_\mu=i\hbar\partial_\mu=i\hbar\frac{\partial}{\partial x^\mu}=(p_0,-{\bf p})$ is the four-momentum operator, and $x^\mu=(ct,{\bf r})$ is the four-position operator (for simplicity we omit the symbol for quantum operators). In particular, here we have basically the DE with a type of nonminimal coupling, given by the second term of Eq. \eqref{5} (another type of nonminimal coupling is for the magnetic dipole moment, for example \cite{BJP,Greiner,Hagen2}). Thus, the EDM is also sometimes called of ``nonminimal coupling constant''. So, based on the fact that $\gamma^\mu p_\mu=\gamma^0\frac{i\hbar}{c}\frac{\partial}{\partial t}-\boldsymbol{\gamma}\cdot{\bf p}$, and $\sigma^{\mu\nu}\gamma_5 F_{\mu\nu}=2\left(\frac{2i}{\hbar}{\bf S}\cdot{\bf E}-c\boldsymbol{\alpha}\cdot{\bf B}\right)$, Eq. \eqref{5} is rewritten in the form (differential DE)
\begin{equation}\label{6}
\left[c\boldsymbol{\alpha}\cdot({\bf p}+id_f\beta{\bf B})-\frac{2d_f}{\hbar}\beta{\bf S}\cdot{\bf E}+\beta m_0 c^2-i\hbar\frac{\partial}{\partial t}\right]\Psi=0,
\end{equation}
or in terms of the total Hamiltonian (operator) of the system $H_T$, as
\begin{equation}\label{7}
i\hbar\frac{\partial\Psi}{\partial t}=H_T\Psi=[H_0+H_{EDM}]\Psi,
\end{equation}
where $H_0=c\boldsymbol{\alpha}\cdot{\bf p}+\beta m_0 c^2$ is the free Dirac Hamiltonian \cite{Greiner}, $H_{EDM}=-d_f \beta\left(\frac{2}{\hbar}{\bf S}\cdot{\bf E}+ic\boldsymbol{\alpha}\cdot{\bf B}\right)$ is the Dirac Hamiltonian for the EDM, or simply the Hamiltonian of the EDM \cite{Commins}, being ${\bf S}=\frac{\hbar}{2}\boldsymbol{\Sigma}$ the spin operator (vector), where $\boldsymbol{\Sigma}=1\otimes\boldsymbol{\sigma}=(\Sigma^1,\Sigma^2,\Sigma^3)$, with $\Sigma^i=\frac{i}{2}\epsilon^{ijk}\gamma_j \gamma_k$, ${\bf E}=-\boldsymbol{\nabla}A_0-\frac{\partial{\bf A}}{\partial t}$ is the electric field, with ${\bf S}\cdot{\bf E}$ being the ``spin-electric field coupling'', and ${\bf B}=\boldsymbol{\nabla}\times {\bf A}$ is the magnetic field.

On the other hand, here we assume that our system is a stationary system, or composed of stationary states (the Hamiltonian does not depend explicitly on time); consequently, the Dirac spinor can be written as follows \cite{Greiner,Bermudez2}
\begin{equation}\label{spinor1}
\Psi=e^{-iEt/\hbar}\psi,
\end{equation}
where $E$ is the relativistic total energy, and $-\infty<t<\infty$ is the temporal coordinate (proper time). Besides, modifying the momentum in the form ${\bf p}\rightarrow {\bf p}-im_{0}\omega\beta {\bf r}$, we then have from \eqref{6} the following (time-independent) stationary DO with EDM
\begin{equation}\label{8}
\left[c\boldsymbol{\alpha}\cdot({\bf p}-im_{0}\omega\beta {\bf r}+id_f\beta{\bf B})-\frac{2d_f}{\hbar}\beta{\bf S}\cdot{\bf E}+\beta m_0 c^2-E\right]\psi=0,
\end{equation}
or in the form of an eigenvalue equation, as
\begin{equation}\label{9}
\bar{H}_{DO}\psi=[H_{DO}+H_{EDM}]\psi=E\psi,
\end{equation}
where $\bar{H}_{DO}$ is the Hamiltonian of the DO with EDM or the total Hamiltonian of the DO, being $H_{DO}=c\boldsymbol{\alpha}\cdot({\bf p}-im_0\beta{\bf r})+\beta m_0 c^2$ the usual DO Hamiltonian, $\psi=\ket{\psi}$ is the eigenfunction or an eigenvector (time-independent Dirac spinor), and $E$ is the eigenvalue (eigenenergy).

Here, we will also consider an intrinsically two-dimensional magnetic field ${\bf B}$ (it has only two spatial components), and also a fermion confined exclusively to the surface (Cartesian or polar plane), which implies $p_z=z=0$ \cite{Andrade2}. However, the electric field can have components at z, but not depend on z, otherwise, we would have no electric field acting on the system. In that way, in (2+1)-dimensions where $\boldsymbol{\alpha}=\boldsymbol{\Sigma}=\boldsymbol{\sigma}$ and $\gamma^0=\sigma_3$ \cite{Andrade1,BJP,Bermudez1,Bermudez2,Villalba}, we have
\begin{equation}\label{10}
\left[c\boldsymbol{\sigma}\cdot({\bf p}-im_{0}\omega\sigma_3{\bf r}+id_f\sigma_3{\bf B})-d_f\sigma_3\boldsymbol{\sigma}\cdot{\bf E}+\sigma_3 m_0 c^2-E\right]\psi=0,
\end{equation}
where ${\bf p}=(p_1,p_2)=(p_x,p_y)$ is the momentum operator, ${\bf r}=(x_1,x_2)=(x,y)$ is the position vector, $\psi$ is now a two-component Dirac spinor (which mixes spin-up and down components with positive and negative energies), and $\boldsymbol{\sigma}=(\sigma_1,s\sigma_2)=(\sigma_x,s\sigma_y)$, where the spin parameter $s$ takes the values $\pm 1$, being $s=+1$ for the spin ``up'' (``$\uparrow$''), and $s=-1$ for the spin ``down'' (``$\downarrow$''), respectively \cite{Andrade1,Andrade2,Villalba,Oliveira2,Oliveira3}. In particular, this parameter was introduced in the literature in 1990 through two articles published by C. R. Hagen (Aharonov-Bohm Scattering of Particles with Spin, and Exact Equivalence of Spin-1/2 Aharonov-Bohm and Aharonov-Casher Effects), and it was a way to include the ``spin'' in planar fermions (both effects are intrinsically planar phenomena) \cite{Hagen1,Hagen2}.

Let's now choose the form (configuration) of the external electromagnetic field. Then, in polar coordinates (but could be in Cartesian coordinates), we consider a radial magnetic field and linear in $r=\sqrt{x^2+y^2}>0$ (radial coordinate) generated by anti-Helmholtz coils, given by: ${\bf B}=\Phi{\bf r}=\Phi r\hat{e}_r$ (also called of anti-Helmholtz magnetic-field), where $\Phi=\frac{B_0}{R}>0$ is a linear density of magnetic field, being $B_0$ the averaged field strength when $r$ is equal to the radius ($r=R$), and $\hat{e}_r$ is a unit vector in the radial direction \cite{Oli,Lira1,Lira2,Livera,Chen1,Chen2}. In particular, such a magnetic field (the general case have a dependency on z) is produced by a pair of identical Helmholtz coils whose electric currents are equal and flow in opposite directions (anti-Helmholtz configuration), and is used in problems involving ferrofluids and magnetorheological fluids \cite{Oli,Lira1,Lira2,Livera,Chen1,Chen2}, magnetic traps for neutral atoms and molecules and two-dimensional magneto-optical traps \cite{Bergeman,Balykin,Harris,Dieckmann}, synthesis of cold antihydrogen atoms \cite{Enomoto,Nagata}, fabrication of two-dimensional electromagnetic scanning micromirrors \cite{Ji}, etc. In addition, it is interesting to comment that radial magnetic fields are also important in astrophysics and solar physics (in solar winds) \cite{Jokipii,Smith,Gosling,Owens}. Now, with respect to the electric field, we consider the uniform electric field of the Stark effect, given by: ${\bf E}=E_0 \hat{e}_z$, where $E_0>0$ is the field strength, and $\hat{e}_z$ is a unit vector in the longitudinal (axial) direction \cite{Sakurai}. In particular, the Stark effect arises due to the interaction of the EDM of atoms or molecules with a uniform external electric field, and there are two types of Stark effect: the linear (for a permanent EDM) and the quadratic (for an induced EDM). In this paper, we have the first case.

With the well-defined electromagnetic field, Eq. \eqref{10} becomes (Cartesian coordinates)
\begin{equation}\label{11}
\left[c\sigma_1(p_x-iAx\sigma_3)+sc\sigma_2(p_y-iAy\sigma_3)-d_f E_0+\sigma_3 m_0 c^2-E\right]\psi=0,
\end{equation}
where $A\equiv(m_{0}\omega-d_f\Phi)>0$, and we have used ${\bf B}=(\Phi x,\Phi y)$.

So, to obtain the (linear) NCDO, it is necessary to write the momentum and position operators, well as the spinor, in a NCPS. Having done that, we have the following NCDO
\begin{equation}\label{12}
\left[c\sigma_1(p_x^{NC}-iAx^{NC}\sigma_3)+sc\sigma_2(p_y^{NC}-iAy^{NC}\sigma_3)-d_f E_0+\sigma_3 m_0 c^2-E\right]\star\psi=0,
\end{equation}
or explicitly, as
\begin{equation}\label{13}
\left[c\lambda\sigma_1(p_x+isp_y\sigma_3)-m_0 c\Omega\sigma_2(x+isy\sigma_3)-\left(E+d_f E_0\right)+\sigma_3 m_0 c^2\right]\psi^{NC}=0,
\end{equation}
where $\lambda\equiv\left(1+\frac{As\theta}{2\hbar}\right)>0$, $\Omega\equiv\left(\frac{A}{m_0}+\frac{s\eta}{2\hbar m_0}\right)>0$, and $\psi^{NC}$ is our NC Dirac spinor, and we use the fact that $p_x^{NC}$ and $x^{NC}$ are given by
\begin{equation}\label{p}
p^{NC}_x=p_x+\frac{\eta}{2\hbar}y, \ \ p^{NC}_y=p_y-\frac{\eta}{2\hbar}x,
\end{equation}
\begin{equation}\label{x}
x^{NC}=x-\frac{\theta}{2\hbar}p_y, \ \ y^{NC}=y+\frac{\theta}{2\hbar}p_x.  
\end{equation}

Now, let's get the quadratic NCDO, given by a second-order differential equation (for the two components of the spinor), where it is from this equation that we will determine the bound-state solutions (or stationary-state solutions) of the problem, given by the Dirac spinor and the relativistic energy spectrum, respectively. For this, we will define an ansatz for the original Dirac spinor according to Refs. \cite{Andrade2,Gavrilov}, which can be defined in terms of another spinor as follow (already included the nonminimal coupling)
\begin{equation}\label{Diracspinor}
\psi^{NC}\equiv O_D\varphi^{NC}=\left[c\lambda\sigma_1(p_x+isp_y\sigma_3)-m_0 c\Omega\sigma_2(x+isy\sigma_3)+\left(E+d_f E_0\right)+\sigma_3 m_0 c^2\right]\varphi^{NC},
\end{equation}
where $O_D$ is a ``Dirac operator'', and has a similar form to the NCDO. From a physical point of view, a spinor in the form \eqref{13} aims to arrive at the expression of the relativistic energy-momentum relation, or relativistic dispersion relation, given by $E^2={\bf p}^2 c^2 +m_0^2 c^4$ (for a relativistic free particle), in which every relativistic wave equation must satisfy \cite{Greiner,Sakurai,Bermudez1}.

Then, substituting \eqref{Diracspinor} in \eqref{13}, we get the following quadratic NCDO
\begin{equation}\label{14}
\left[c^2\lambda^2(p_x^2+p_y^2)+m_0^2 c^2\Omega^2(x^2+y^2)-2m_0c^2\lambda\Omega(\hbar\sigma_3+sL_z)-(E+d_f E_0)^2+m_0^2 c^4\right]\varphi^{NC}=0,
\end{equation}
or in the form of an eigenvalue equation, as
\begin{equation}\label{15}
H_{NCDO}^{quadratic}\varphi^{NC}=\left[H^{2D}_{NCQHO-like}-\frac{\Omega}{\lambda}(\hbar\sigma_3+sL_z)\right]\varphi^{NC}=\left[\frac{(E+d_f E_0)^2-m_0^2 c^4}{2m_0 c^2 \lambda^2}\right]\varphi^{NC},
\end{equation}
where 
\begin{equation}\label{QHO}
H^{2D}_{NCQHO-like}=\frac{{\bf p}^2}{2m_0}+\frac{1}{2}\left(\frac{m_0\Omega^2}{\lambda^2}\right){\bf r}^2,
\end{equation}
with $H_{NCDO}^{quadratic}$ being the Hamiltonian of the quadratic NCDO, $H^{2D}_{NCQHO-like}$ is the noncommutative quantum harmonic oscillator (NCQHO)-like Hamiltonian in two-dimensions (since $\lambda$ is dimensionless and $\Omega$ has angular frequency dimension), and ${\bf p}$, ${\bf r}$, and $L_z$ are given by ${\bf p}=(p_x^2+p_y^2)$, ${\bf r}=x^2+y^2$, and $L_z=xp_y-yp_x$ (z-component of the orbital angular momentum operator ${\bf L}$). It is worth noting that in the absence of the EDM ($d_f=0$), or of the electromagnetic field (${\bf B}={\bf E}=0$), and of the NCPS ($\theta=\eta=0$), with $\varphi=[\ket{\psi_1},\ket{\psi_2}]^T \in \mathbb{C}^2$, we reduce Eq. \eqref{15} to the particular case of literature (DO in the position space) \cite{Andrade1,Bermudez2,Rao}. 

On the other hand, since we want to obtain the bound-state solutions for the NCDO in polar coordinates, we must write ${\bf p}$, ${\bf r}$, and $L_s$ as follow
\begin{equation}\label{16}
{\bf p}=-i\hbar\left(\hat{e}_r\frac{\partial}{\partial r}+\frac{\hat{e}_\phi}{r}\frac{\partial}{\partial\phi}\right), \ \ {\bf r}=r\hat{e}_r, \ \ L_z=-i\hbar\frac{\partial}{\partial\phi},
\end{equation}
where $0\leq\phi\leq 2\pi$ is the angular coordinate and $r=\sqrt{x^2+y^2}$, with $0\leq r<\infty$, is the polar radial coordinate, respectively. In addition, a good choice for the spinor $\varphi^{NC}$ (also in polar coordinates) can be write as follow \cite{Andrade1,Andrade2}
\begin{equation}\label{spinor2}
\varphi^{NC}(r,\phi)=\left(
           \begin{array}{c}
            e^{im\phi}f_+(r) \\
            e^{i(m+s)\phi}f_-(r) \\
           \end{array}
         \right), \ \ (f_+(r)\neq f_-(r)),
\end{equation}
where $f_\pm(r)$ are real radial functions (spinorial radial components), and $m=0,\pm 1,\pm 2, \ldots$ is the orbital magnetic quantum number (or simply magnetic quantum number). Here, we wrote our spinor in the form \eqref{spinor2} to be in agreement with the particular case of literature (as we will see shortly). In addition, a spinor in the form $\varphi^{NC}(r,\phi)=e^{im\phi}[f_+(r),f_-(r)]^T$ it would no longer be consistent, since that would imply $f_+(r)=f_-(r)$ as a solution of a  (same) differential equation, what is not allowed. Otherwise, it would be the same as saying that a particle is equal to its own antiparticle, which is not the case since we are dealing with Dirac fermions and not Marjorana fermions \cite{Greiner}.

Therefore, substituting \eqref{spinor2} and \eqref{16} into \eqref{15}, we get the following second-order differential equation for the NCDO (radial or quadratic NCDO)
\begin{equation}\label{17}
\left[\frac{d^2}{dr^2}+\frac{1}{r}\frac{d}{dr}-\frac{\Gamma^2_\kappa}{r^2}-\Lambda^2r^2+E_\kappa\right]f_\kappa(r)=0,
\end{equation}
where we define
\begin{equation}\label{18}
\Gamma_\kappa\equiv\left(m+s\frac{1-\kappa}{2}\right), \ \ E_\kappa\equiv\frac{(E+d_f E_0)^2-m_0^2c^4}{(\hbar c \lambda)^2}+2s\Lambda(\Gamma_\kappa+s\kappa), \ \ \Lambda\equiv\frac{m_0\Omega}{\hbar\lambda},
\end{equation}
with $\kappa=\pm 1$ being a parameter that describes the two components of the spinor, where $\kappa=+1$ describes a ``particle with spin up or down'' ($s=\pm 1$), and $\kappa=+1$ describes an ``antiparticle with spin up or down'' ($s=\pm 1$), respectively. So, it is worth mentioning that in the absence of the EDM ($d_f=0$), or of the electromagnetic field (${\bf B}={\bf E}=0$), and also of the NCPS ($\theta=\eta=0$), we reduce Eq. \eqref{17} to the particular case of literature \cite{Andrade1}. Already for $d_f=0$, and without the ``spin'' ($s=+1$), with $x=i\hbar\frac{\partial}{\partial p_x}$, $y=i\hbar\frac{\partial}{\partial p_y}$, we reduce Eq. \eqref{17} (for $\kappa=+1$) also to a particular case of literature (NCDO in the momentum space) \cite{Boumali2}.

\section{Bound-state solutions: two-component Dirac spinor and the relativistic energy spectrum}\label{sec4}

To solve exactly Eq. \eqref{17}, let's introduce a new (dimensionless) variable in our problem, given by: $\rho=\Lambda r^2$ ($\Lambda=\vert\Lambda\vert>0$). Thus, through a change of variable, Eq. \eqref{17} becomes
\begin{equation}\label{19}
\left[\rho\frac{d^{2}}{d\rho^{2}}+\frac{d}{d\rho}-\frac{\Gamma^{2}_\kappa}{4\rho}-\frac{\rho}{4}+\frac{E_\kappa}{4\Lambda}\right]f_\kappa(\rho)=0.
\end{equation}

Now, analyzing the asymptotic (limit) behavior of Eq. (\ref{19}) for $\rho\to 0$ and $\rho\to\infty$, we have a (regular) solution to this equation given by the following ansatz
\begin{equation}\label{20}
f_\kappa(\rho)=C_\kappa\rho^{\frac{\vert\Gamma_\kappa\vert}{2}}e^{-\frac{\rho}{2}}F_\kappa(\rho), \ \ (\vert\Gamma_\kappa\vert\geq 0),
\end{equation}
where $C_\kappa>0$ are normalization constants, $F_\kappa(\rho)$ are unknown functions to be determined, and $f_\kappa(\rho)$ must satisfy the following boundary conditions to be a normalizable solution
\begin{equation}\label{conditions} 
f_\kappa(\rho\to 0)=f_\kappa(\rho\to\infty)=0.
\end{equation}

Substituting \eqref{20} in \eqref{19}, we have a second-order differential equation for $F_\kappa(\rho)$ as follows
\begin{equation}\label{21}
\left[\rho\frac{d^{2}}{d\rho^{2}}+(\vert\bar{\Gamma}_\kappa\vert-\rho)\frac{d}{d\rho}-\bar{E}_\kappa\right]F_\kappa(\rho)=0,
\end{equation}
where
\begin{equation}\label{22}
\vert\bar{\Gamma}_\kappa\vert\equiv\vert\Gamma_\kappa\vert+1, \ \ \bar{E}_\kappa\equiv\frac{\vert\bar{\Gamma}_\kappa\vert}{2}-\frac{E_\kappa}{4\Lambda}.
\end{equation}

According to the literature \cite{Andrade1,Villalba}, Eq. \eqref{21} is the well-known generalized Laguerre equation, whose solution are the so-called generalized Laguerre polynomials, written as $F_\kappa(\rho)=L^{\vert\Gamma_\kappa\vert}_n(\rho)$. Consequently, the quantity $\bar{E}_\kappa$ must to be equal to a non-positive integer, given by: $\bar{E}_\kappa=-n$, where $n=n_r=0,1,2,\ldots$ is the radial quantum number (because it arises directly from the radial NCDO), and represents the number of nodes of the radial functions $f_\kappa (r)$. Therefore, from this quantization condition ($\bar{E}_\kappa=-n$), we obtain as a result the following relativistic energy spectrum (relativistic energy levels) for the NCDO with EDM in the presence of an external electromagnetic field
\begin{equation}\label{spectrum}
E^{\chi}_{n,m,s}=U+\chi m_0c^2\sqrt{1+\frac{2\hbar N}{m_0c^2}\left(1+s\frac{(\omega-\sigma\tilde{\omega})}{\omega_\theta}\right)(\omega+s\omega_\eta-\sigma\tilde{\omega})},
\end{equation}
where
\begin{equation}\label{potentialenergy}
U\equiv-d_f E_0=-\sigma\vert d_f\vert E_0, \ \ (\sigma=\pm 1),
\end{equation}
and
\begin{equation}\label{N}
N=N_{eff}\equiv\left[2n+1-\kappa+\Big|m+s\frac{1-\kappa}{2}\Big|-s\left(m+s\frac{1-\kappa}{2}\right)\right]\geq 0,
\end{equation}
being $\chi=\pm 1$ a parameter (an ``energy'' parameter) in which it describes the positive energy states ($\chi=+1$), or simply a particle (neutron or DO), as well as the negative energy states ($\chi=-1$), or simply an antiparticle (antineutron or anti-DO), and $N$ is a total or effective quantum number (depends on everyone else). Then, we clearly see that the spectrum \eqref{spectrum} is a linear function on the (electric) potential energy $U$ (an energy generated by the interaction of the EDM with the external electric field ${\bf E}$) \cite{Jackson,Lacava}, where $\sigma=+1$ is for $d_f>0$ and $\sigma=-1$ is for $d_f<0$, and also depends explicitly on the quantum numbers $n$ and $m$, spin parameter $s$, and of four angular frequencies, given by: $\omega$, $\tilde{\omega}$, $\omega_{\theta}$, and $\omega_{\eta}$, where $\omega$ is the well-known frequency of the NCDO, $\tilde{\omega}\equiv\frac{\vert d_f \vert \Phi}{m_0}>0$ is a type of ``cyclotron frequency'' (in this case $\sigma=\pm 1$ describes the revolution direction of the corresponding classical motion), and $\omega_{\theta}\equiv\frac{2\hbar}{m_0\theta}>0$ and $\omega_{\eta}\equiv\frac{\eta}{2\hbar m_0}>0$ are the NC frequencies of position and momentum, respectively. Therefore, we can say that the first term of \eqref{spectrum} is the non-quantized (continuous) part of the spectrum (unaffected by NC parameters) and the second term is the quantized (discrete) part (affected by NC parameters). In addition, as a direct consequence of energy $U$, the spectrum becomes asymmetric (there is a ``break'' of symmetry in the spectrum), i.e., the energies of the particle and antiparticle are not equal ($E^+\neq\vert E^-\vert$). Still on the energy $U$, and making an analogy with the classical case (a classical EDM), $U$ can be positive: $U>0$ (the energy is highest), in which means an EDM antiparallel to the electric field, or negative: $U<0$ (the energy is lowest), in which means an EDM parallel to the electric field, respectively. On the other hand, we note that even in the absence of the DO ($\omega=0$) and of the EDM ($d_f=0$), the spectrum still remains quantized due to the presence of $\omega_{\eta}$ or $\eta$ (not of $\omega_{\theta}$), in which such frequency or parameter acts as a kind of ``NC field'', and whose spectrum is given by:
\begin{equation}\label{NCspectrum}
E^{\chi}_{n,m,s}=\chi m_0 c^2\sqrt{1+\frac{2\hbar\omega_\eta}{m_0 c^2}\bar{N}}, 
\end{equation}
where $\bar{N}\equiv 2n+1-s\kappa+\big|m+s\frac{1-\kappa}{2}\big|-\left(m+s\frac{1-\kappa}{2}\right)$ (once now $\Lambda=s\omega_\eta$, and $\Lambda\neq\vert\Lambda\vert$), in which is the relativistic spectrum for a ``free'' Dirac fermion in a NCPS with the NC only of the momentum (not yet studied in the literature). Already in the absence of the DO ($\omega=0$) and of the NCPS ($\theta=\eta=0$), the spectrum also remains quantized (due to the presence of $\tilde{\omega}$ or $d_f$), and written as
\begin{equation}\label{EDMspectrum}
E^{\chi}_{n,m,s}=U+\chi m_0 c^2\sqrt{1+\frac{2\hbar\tilde{\omega}}{m_0 c^2}\tilde{N}}, 
\end{equation}
where $\tilde{N}\equiv 2n+1+\sigma\kappa+\big|m+s\frac{1-\kappa}{2}\big|+s\left(m+s\frac{1-\kappa}{2}\right)$ (once now $\Lambda=-\sigma\frac{m_0\tilde{\omega}}{\hbar}$, and $\Lambda\neq\vert\Lambda\vert$), in which is the relativistic spectrum of a ``free'' neutral Dirac fermion with EDM (not yet studied in the literature).

Furthermore, it is also interesting to note that the conditions for $\lambda$ and $\Omega$, which in the spectrum \eqref{spectrum} are given by the following expressions
\begin{equation}\label{conditions1}
\lambda=\left(1+s\frac{(\omega-\sigma\tilde{\omega})}{\omega_\theta}\right)>0,
\end{equation}
and
\begin{equation}\label{conditions2}
\Omega=(\omega+s\omega_\eta-\sigma\tilde{\omega})>0,
\end{equation}
results in possible values at which $\theta$ (or $\omega_\theta$) and $\eta$ (or $\omega_\eta$) must satisfy, and are given as follows
\begin{equation}\label{conditions3}
\theta>\frac{2s\hbar}{(\sigma\tilde{\omega}-\omega)}, \ \ or \ \ \omega_\theta<s(\sigma\tilde{\omega}-\omega), \ \ (s^2=1),
\end{equation}
and
\begin{equation}\label{conditions3}
\eta>2s\hbar m_0(\sigma\tilde{\omega}-\omega), \ \ or \ \ \omega_\eta>s(\sigma\tilde{\omega}-\omega), \ \ (s^2=1).
\end{equation}

Before we analyze graphically (via 2D graphs) and in detail the behavior of the spectrum \eqref{spectrum} as a function of the four angular frequencies for different values of $n$ and $m$, with and without the influence of $U$ (or $E_0$), it is interesting that we first analyze one of the most important and striking aspects of two-dimensional energy spectra (or three-dimensional) \cite{Moshinsky,Andrade1,Strange}, which is their degeneracy or the degenerate states of the system. In that way, we verify that this degeneracy depends on the values (signs) of $s$ and $m$, however, is not affected by the parameter $\kappa$ (or $\sigma$). Therefore, in Table \eqref{tab1} we have four possible configurations for the degeneracy depending on the values of $s$ and $m$ as well as the respective values of $N$.
\begin{table}[h]
\centering
\begin{small}
\caption{Degeneracy depends on the values of $s$ and $m$.} \label{tab1}
\begin{tabular}{ccccc}
\hline
Configuration & $s$ & $m$ & $N$ & Degeneracy \\
\hline
1& $+1$ & $m\geq 0$ & $2n+1-\kappa$ & infinite\\
2& $+1$ & $m<0$ & $2n+1-\kappa+2\vert m+\frac{1-\kappa}{2}\vert$ & finite \\
3& $-1$ & $m\geq 0$ & $2n+1-\kappa+2\vert m-\frac{1-\kappa}{2}\vert$ & finite\\
4& $-1$ & $m<0$ & $2n+1-\kappa$ &  infinite\\
\hline
\end{tabular}
\end{small}
\end{table}

According to Table \eqref{tab1}, we see that the spectrum can be infinitely (for $sm\geq 0$) or finitely (for $sm<0$) degenerate, and therefore, there can be a finite or infinite number of degenerate states (states that share the same energy eigenvalue) depending on the values of $s$ and $m$ \cite{Moshinsky,Andrade1,Villalba,Pacheco2}. In particular, an infinite (accidental) degeneracy arises when the spectrum (all the energy levels) only depends on the quantum number $n$ (there are infinitely degenerate states), while a finite degeneracy arises when the spectrum (all the energy levels) depends on both the quantum numbers $n$ and $m$ (there are finitely degenerate states) \cite{Moshinsky,Andrade1,Villalba,Pacheco2}. For example, fixing a given value of $\kappa$ (preferably $\kappa=+1$), we can define a new quantum number from $n$ and $m$, such as: $l\equiv n+\vert m\vert\geq 1$ ($\vert m\vert\geq 1$), or $l\equiv n+m\geq 0$ ($m\geq 0$). In that way, from the number $l$ we can determine the expression for the total degree of (finite) degeneracy, given by: $\Omega(l)=\sum\limits_{\vert m\vert=1}^{l}(2\vert m\vert+1)=l(l+2)$, or $\Omega(l)=\sum\limits_{m=0}^{l}(2m+1)=(l+1)^2$, where $2\vert m\vert+1$ ($m<0$ or $m\geq 0$) is the (finite) number of degenerate states of the system.

So, based on the informations about the degeneracy we get Table \eqref{tab2}, where we have four possible configurations for the spectrum depending on the values of $s$ and $m$, in which we define for simplicity that $B\equiv\frac{4\hbar}{m_0c^2}$ and $\tilde{n}\equiv(n+\frac{1-\kappa}{2})$.
\begin{table}[h]
\centering
\begin{small}
\caption{Relativistic energy spectra for the degenerate states of the particle and antiparticle.} \label{tab2}
\begin{tabular}{ccc}
\hline
Configuration & Relativistic energy spectrum & Degeneracy \\
\hline
1 &$E^{\chi}_{n,+}=U+\chi m_0c^2\sqrt{1+\tilde{n}B\left(1+\frac{(\omega-\sigma\tilde{\omega})}{\omega_\theta}\right)(\omega+\omega_\eta-\sigma\tilde{\omega})}$ & infinite \\
2 &$E^{\chi}_{n,m<0,+}=U+\chi m_0c^2\sqrt{1+B\left(1+\frac{(\omega-\sigma\tilde{\omega})}{\omega_\theta}\right)(\omega+\omega_\eta-\sigma\tilde{\omega})(\tilde{n}+\vert m+\frac{1-\kappa}{2}\vert)}$ & finite \\
3 &$E^{\chi}_{n,m\geq 0,-}=U+\chi m_0c^2\sqrt{1+B\left(1-\frac{(\omega-\sigma\tilde{\omega})}{\omega_\theta}\right)(\omega-\omega_\eta-\sigma\tilde{\omega})(\tilde{n}+\vert m-\frac{1-\kappa}{2}\vert)}$ & finite \\
4 &$E^{\chi}_{n,-}=U+\chi m_0c^2\sqrt{1+\tilde{n}B\left(1-\frac{(\omega-\sigma\tilde{\omega})}{\omega_\theta}\right)(\omega-\omega_\eta-\sigma\tilde{\omega})}$ & infinite \\
\hline
\end{tabular}
\end{small}
\end{table}

According to Table \eqref{tab2}, we see that for $sm\geq 0$ (configs. 1 and 4), the spectra increase as a function of $n$, while for $sm<0$ (configs. 2 and 3), the spectra increase as a function of $n$ and $m$, respectively. With respect to the maximal and minimal spectrum, the maximal spectrum (highest allowed energies) is reached for $s=+1$ and $m<0$ with $\sigma=-1$ and $\kappa=\pm 1$ (config. 2), while the minimal spectrum (lowest allowed energies) is reached for $s=-1$ and $m<0$ with $\sigma=\kappa=+1$ (config. 4). Furthermore, when $U\to 0$ (without electric field) all spectra from Table \eqref{tab2} become symmetrical (for each specific configuration), and therefore, we have: $E^+=\vert E^-\vert>0$. On the other hand, comparing the spectrum \eqref{spectrum}, or the spectra from Table \eqref{tab2}, with some works of the literature, we verified that in the absence of the EDM ($d_f=0$), or of the electromagnetic field (${\bf B}={\bf E}=0$), and of the NCPS ($\theta=\eta=0$), we obtain the usual spectrum of the DO for $\kappa=+1$ with $sm\geq 0$ or $sm<0$ \cite{Andrade1,Villalba,Bermudez1,Bermudez2}, and for $\kappa=-1$ with $sm\geq 0$ \cite{Rao}. Already for $d_f=0$ with $\kappa=+1$ and $sm\geq 0$, we obtain the spectrum of the NCDO \cite{Boumali2}. It is interesting to note that for $d_f=\theta=\eta=0$ with $\kappa=+1$ and $sm\geq 0$, we obtain also the spectrum of the DO in (3+1)-dimensions for $j=l+1/2=1/2,3/2,\ldots$ $(E^{(2+1)D}_{n,ms\geq 0,\kappa=+1}=E^{(3+1)D}_{n,j=l+1/2})$, where $j$ is the total angular momentum quantum number that arises from ${\bf J}^2$ (total angular momentum squared), $l$ is the orbital (azimuthal) quantum number that arises from ${\bf L}^2$ (orbital angular momentum squared), and $1/2$ is the spin up \cite{Moshinsky,Strange,Szmytkowski,Pacheco2}. However, the spectrum for $j=l-1/2=1/2,3/2,\ldots$ (spin down) is larger for the (3+1)-dimensional case $(E^{(2+1)D}_{n,ms\geq 0,\kappa=+1}<E^{(3+1)D}_{n,j=l-1/2}$ and $E^{(2+1)D}_{n,ms<0,\kappa=+1}<E^{(3+1)D}_{n,j=l-1/2})$, as it should be. So, for $j=l+1/2$ and $sm\geq 0$ the DO does not distinguish a (2+1)-dimensional universe from a (3+1)-dimensional (an ``anomaly''). From the above, we clearly see that our relativistic spectrum generalizes several particular cases of the literature.

\begin{figure}[!h]
\centering
\includegraphics[width=10cm]{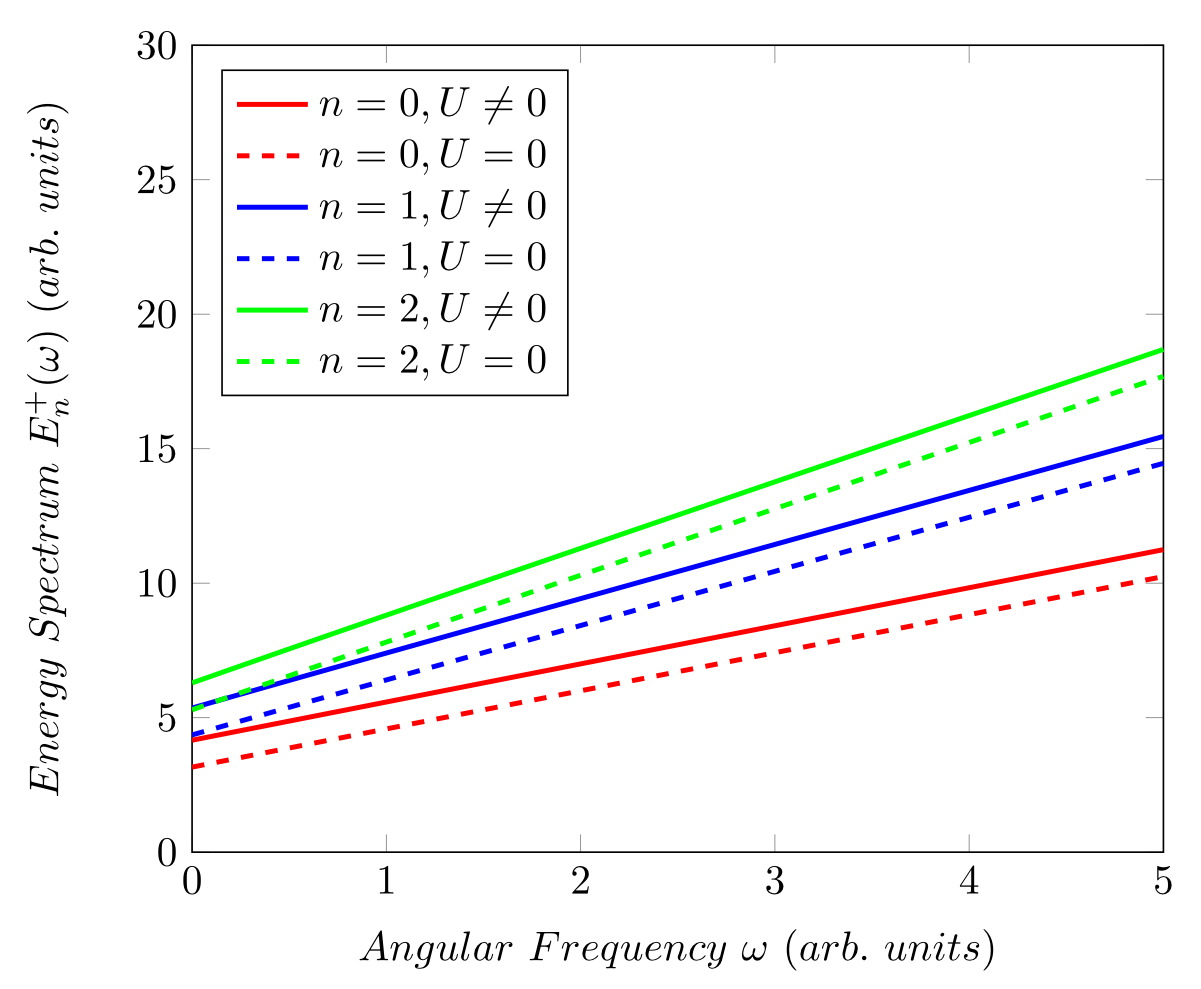}
\caption{Graph of $E^+_n(\omega)$ versus $\omega$ for three different values of $n$ with $U\neq 0$ ($E_0=1$) and $U=0$ ($E_0=0$).}
\label{fig1}
\end{figure}

Now, we can analyze graphically the behavior of the relativistic spectrum as a function of the four angular frequencies for different values of $n$ and $m$ (without loss of generality, here we will adopt only $m=-1$). Actually, such analysis will only be for three frequencies, because as we will see soon, $\omega$ and $\tilde{\omega}$ have the same behavior (for $\sigma=-1$). Besides, for the sake of practicality, we will focus our attention, for example, on the maximal spectrum (config. 2). Therefore, first considering the particle, we have Fig. \eqref{fig1}, where it shows the behavior of the energies as a function of $\omega$ for the ground state ($n=0$) and the first two excited states ($n=1,2$), with and without the presence of potential energy $U$ ($U\geq 0$), in which we take for simplicity that $\hbar=c=m_0=\vert d_f\vert=\Phi=\theta=\eta=1$.

According to Fig. \eqref{fig1}, we see that the energies increase linearly as a function of $\omega$ (the higher $\omega$ the higher $E^+_n (\omega)$), and their values are larger with the increase of $n$ (as it should be) and in the presence of $U$. Therefore, the function of the electric field $E_0$ is to increase the energies of the particle (and also increases linearly as a function of $E_0$ if the quantized part remains constant). In addition, the graph of $E^+_n(\tilde{\omega})$ versus $\tilde{\omega}$ is identical to Fig. \eqref{fig1}, and as $\tilde{\omega}\propto B_0$, implies that the energies increase linearly as a function of the magnetic field. In fact, the behavior between both graphs happens because these two frequencies are ``indistinguishable or interchangeable variables'' ($\omega\leftrightarrow\tilde{\omega}$), that is, we can exchange one frequency for the other and the graphs will be the same (however, this is only true for $d_f<0$ or $\sigma=-1$, or even if $d_f>0$ and ${\bf B}<0$). Thus, we can give another interpretation (another possible explanation) for the emergence of the nonminimal coupling of the DO, which would be through a neutral fermion with (negative) EDM interacting with a radial magnetic field and linear at $r$, where $d_f{\bf B}\to -m_0\omega{\bf r}$ (please see Eq. \eqref{8}). From a physical point of view, this interpretation would be the magnetic analog (a ``dual effect'' or ``duality transformation'') of the current nonminimal coupling (neutral fermion with MDM interacting with a radial electric field and linear at $r$), where $\mu{\bf E}\to -m_0 c^2\omega{\bf r}$ ($\mu>0$ and ${\bf E}<0$) \cite{Bentez,Martinez,Moreno,BJP}. It is also worth mentioning that as $\omega$ increases the energy levels become more spaced out (either for the case $U\neq 0$ or $U=0$), i.e., the energy difference between two consecutive levels gets bigger ($\bigtriangleup E_n$ increases).

\begin{figure}[!h]
\centering
\includegraphics[width=10cm]{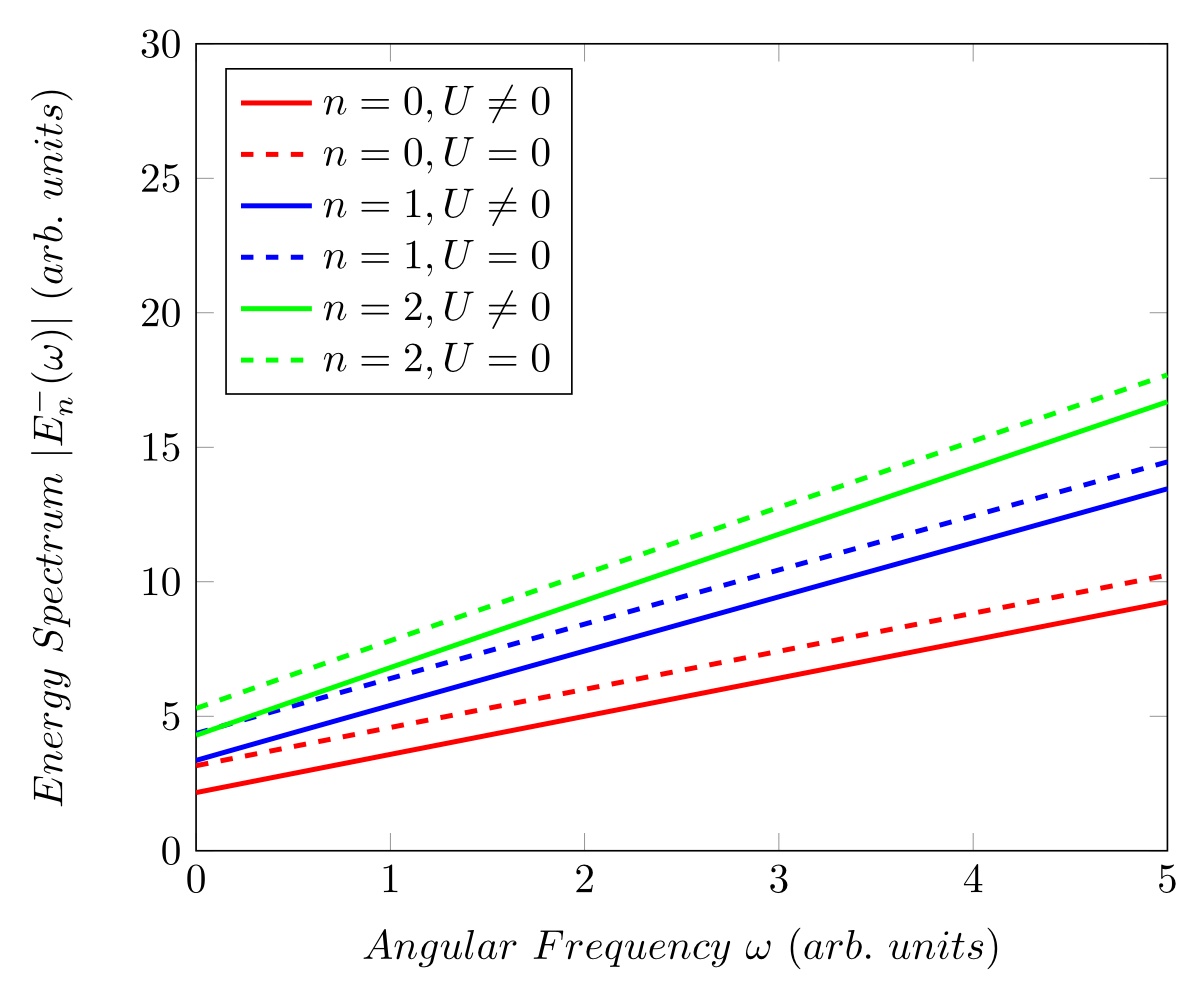}
\caption{Graph of $\vert E^-_n(\omega)\vert$ versus $\omega$ for three different values of $n$ with $U\neq 0$ ($E_0=1$) and $U=0$ ($E_0=0$).}
\label{fig2}
\end{figure}

In Fig. \eqref{fig2}, we have the behavior of the energies of the antiparticle as a function of $\omega$ for the ground state and the first two excited states (with $\hbar=c=m_0=\vert d_f\vert=\Phi=\theta=\eta=1$). According to this figure, we see that the energies increase linearly as a function of $\omega$ (the higher $\omega$ the higher $\vert E^-_n (\omega)\vert$), and their values are larger with the increase of $n$ (as it should be), however, their values are smaller in the presence of $U$. Therefore, the function of the electric field $E_0$ is to decrease the energies of the antiparticle (and decreases linearly as a function of $E_0$ if the quantized part remains constant). So, comparing now both the energies of the particle and antiparticle, we see that the energies of the particle (its energetic content) are always greater than those of the antiparticle for $U\neq 0$, and equal for $U=0$ (symmetrical spectra). Besides, analogous to the case of the particle here the energy levels also become more spaced out. 

\begin{figure}[!h]
\centering
\includegraphics[width=10cm]{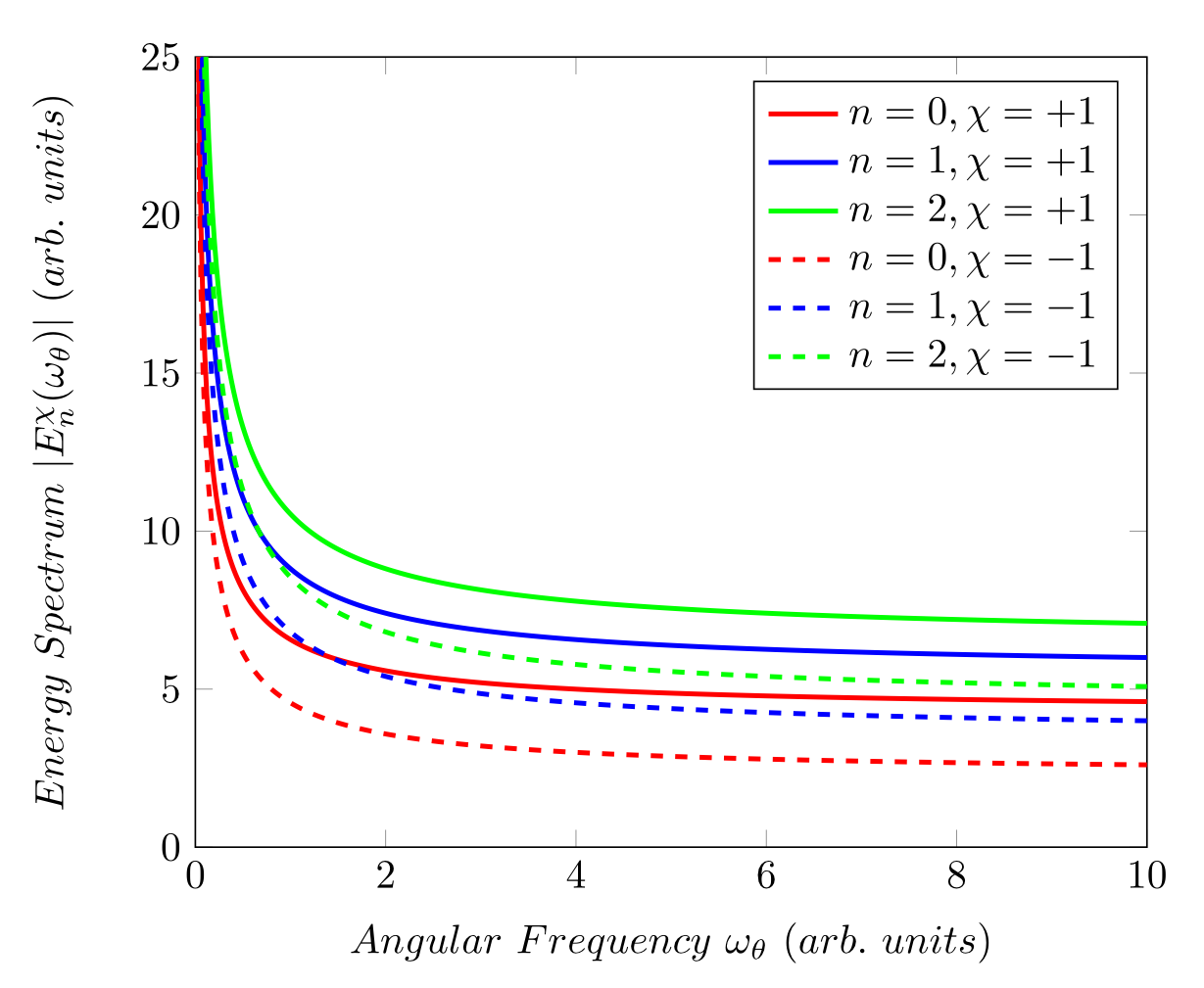}
\caption{Graph of $\vert E^\chi_n(\omega_\theta)\vert$ versus $\omega_\theta$ for three different values of $n$.}
\label{fig3}
\end{figure}

In Fig. \eqref{fig3}, we have the behavior of the energies of the particle ($\chi=+1$) and antiparticle ($\chi=-1$) as a function of $\omega_\theta$ for three different values of $n$ ($n=0,1,2$), where $\hbar=c=m_0=\Phi=\omega=\eta=U=1$. According to this figure, we see that the energies decrease as a function of $\omega_\theta$, however, increase as a function of $\theta$, since $\omega_\theta\propto\frac{1}{\theta}$. Therefore, the function of $\omega_\theta$ is to decrease the energies (the bigger $\omega_\theta$ the smaller $\vert E^\chi_n(\omega_\theta)\vert$), while the function of $\theta$ is to increase (the bigger $\theta$ the bigger $\vert E^\chi_n(\omega_\theta)\vert$). By way of illustration, in the limit $\omega_\theta\to 0$ ($\theta\to\infty$) the energies increase infinitely, while on the limit $\omega_\theta\to\infty$ ($\theta\to 0$) the energies tend to constant values, respectively. Furthermore, we also see that the energies are larger with the increase of $n$ (as it should be). So, comparing both the energies of the particle and antiparticle, we see that the energies of the particle are always greater than those of the antiparticle, except when $\omega_\theta\to 0$ (or $\theta\to\infty$), which is when energy levels get close enough ($\bigtriangleup E\simeq 0$). Now, unlike the graph in Fig. \eqref{fig1} or \eqref{fig2}, here the energy levels have a nearly constant spacing as $\omega_\theta$ increases ($\bigtriangleup E\simeq const.$).

\begin{figure}[!h]
\centering
\includegraphics[width=10cm]{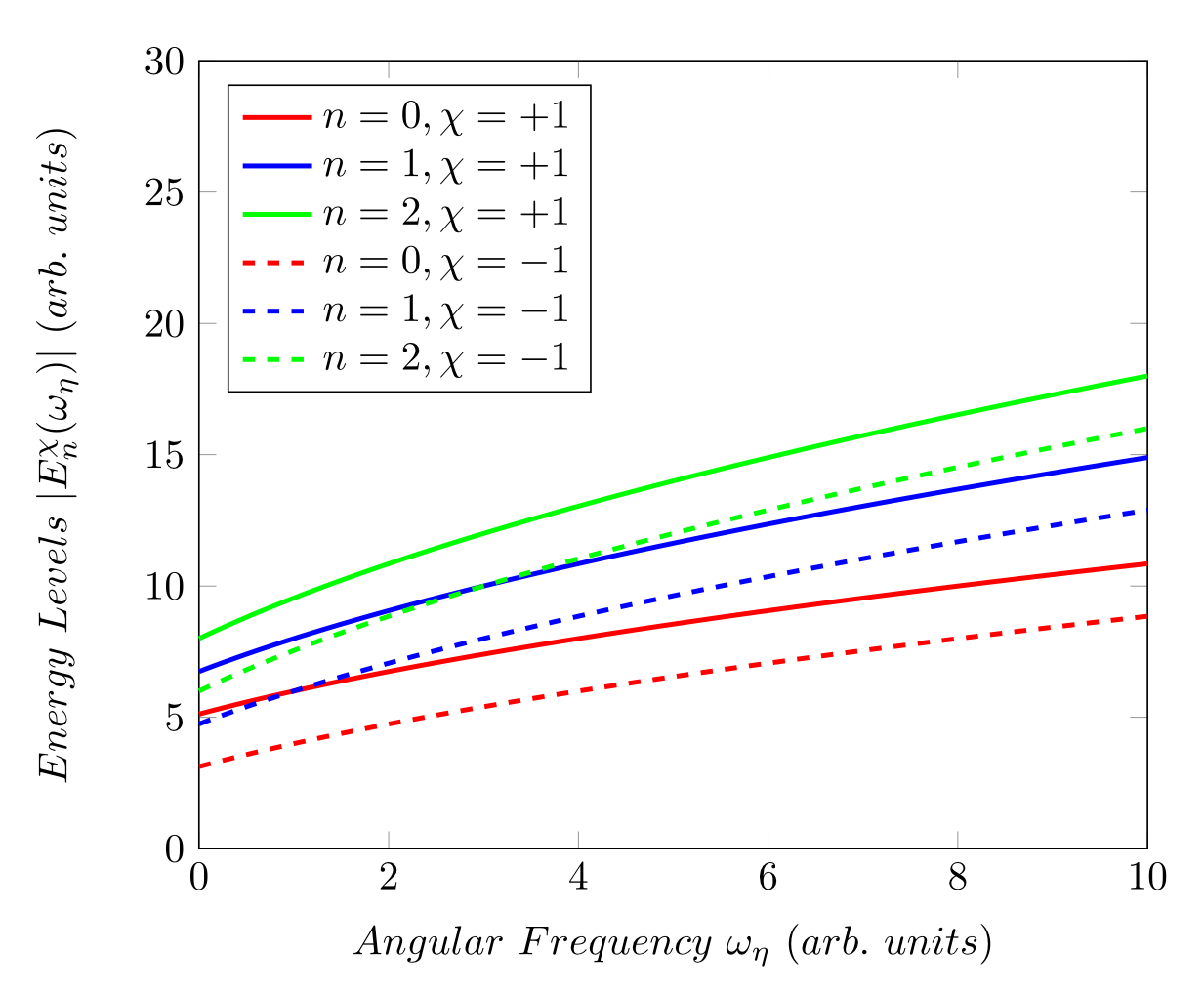}
\caption{Graph of $\vert E^\chi_n(\omega_\eta)\vert$ versus $\omega_\eta$ for three different values of $n$.}
\label{fig4}
\end{figure}

Already in Fig. \eqref{fig4}, we have the behavior of the energies of the particle ($\chi=+1$) and antiparticle ($\chi=-1$) as a function of $\omega_\eta$ for three different values of $n$ ($n=0,1,2$), where $\hbar=c=m_0=\Phi=\omega=\theta=U=1$. According to this figure, we see that the energies increase as a function of $\omega_\eta$, and also of $\eta$, since $\omega_\eta\propto\eta$. Therefore, the function of $\omega_\eta$ (or $\eta$) is to increase the energies (the bigger $\omega_\eta$ or $\eta$ the bigger $\vert E^\chi_n(\omega_\eta)\vert$). Furthermore, we also see that the energies are larger with the increase of $n$ (as it should be). So, comparing both the energies of the particle and antiparticle, we see that the energies of the particle are always greater than those of the antiparticle (whether for large or small values of $\omega_\eta$ or $\eta$). Besides, analogous to the graph in Fig. \eqref{fig1} or \eqref{fig2}, here the energy levels also become more spaced out as $\omega_\eta$ increases ($\bigtriangleup E$ increases).

Now, let's concentrate our attention on the form of the NC Dirac spinor to the relativistic bound states. To obtain such a spinor, we must first find the form of $\psi^{NC}$. So, using the fact that the variable $\rho$ is written as $\rho=\Lambda r^2$, implies that we can rewrite \eqref{20} as
\begin{equation}\label{23}
f_\kappa(r)=\bar{C}_\kappa r^{\vert\Gamma_\kappa\vert}e^{-\frac{\Lambda r^2}{2}}L^{\vert\Gamma_\kappa\vert}_n(\Lambda r^2),
\end{equation}
where $\bar{C}_\kappa\equiv C_\kappa\Lambda^{\frac{\vert\Gamma_\kappa\vert}{2}}>0$ are new normalization constants.

From the radial functions \eqref{23}, the spinor \eqref{spinor2} takes the following form
\begin{equation}\label{spinor3}
\varphi^{NC}(r,\phi)=\left(
           \begin{array}{c}
            \bar{C}_+ e^{im\phi}r^{\vert m\vert}e^{-\frac{\Lambda r^2}{2}}L^{\vert m\vert}_n(\Lambda r^2) \\
            \bar{C}_- e^{i(m+s)\phi}r^{\vert m+s\vert}e^{-\frac{\Lambda r^2}{2}}L^{\vert m+s\vert}_n(\Lambda r^2) \\
           \end{array}
         \right).
\end{equation}

Now, let's find the form of the spinor $\psi^{NC}$. However, we need to write such a spinor in polar coordinates, where $\phi=$arctan$\left(\frac{y}{x}\right)$, and $(x,y)=r(\cos\phi,\sin\phi)$. In that way, using
\begin{equation}\label{24} 
p_x+is\sigma_3 p_y=-i\hbar e^{is\sigma_3 \phi}\left(\frac{\partial}{\partial r}+i\sigma_3\frac{s}{r}\frac{\partial}{\partial\phi}\right), \ \ x+is\sigma_3y=re^{is\sigma_3 \phi},
\end{equation}
the spinor $\psi^{NC}$ becomes
\begin{equation}\label{spinor4}
\psi^{NC}\equiv\left[-i\hbar c\lambda\sigma_1 e^{is\sigma_3 \phi}\left(\frac{\partial}{\partial r}+i\sigma_3\frac{s}{r}\frac{\partial}{\partial\phi}\right)-m_0 c\Omega r\sigma_2e^{is\sigma_3 \phi}+\left(E+d_f E_0\right)+\sigma_3 m_0 c^2\right]\varphi^{NC},
\end{equation}
or explicitly, as
\begin{equation}\label{spinor5}
\psi^{NC}\equiv\left(
    \begin{array}{cc}
      E+d_f E_0+m_0 c^2 & \ -i\hbar c\lambda e^{is\phi}\left(\frac{\partial}{\partial r}-i\frac{s}{r}\frac{\partial}{\partial\phi}\right)+im_0 c\Omega re^{is\phi} \\
      -i\hbar c\lambda e^{is\phi}\left(\frac{\partial}{\partial r}+i\frac{s}{r}\frac{\partial}{\partial\phi}\right)-im_0 c\Omega re^{is\phi} & E+d_f E_0-m_0 c^2 \\
    \end{array}
  \right)\varphi^{NC},
\end{equation}
where we use the fact that $e^{is\sigma_3\phi}=$diag$(e^{is\phi},e^{-is\phi})$, and also the Pauli matrices.

Therefore, replacing \eqref{spinor3} in \eqref{spinor5}, and knowing that $\Psi^{NC}_D=e^{-iEt/\hbar}\psi^{NC}$, we obtain the following two-component NC Dirac spinor
\begin{equation}\label{spinor6} 
\Psi^{NC}_{Dirac}(t,r,\phi)=\left(
           \begin{array}{c}
            e^{i[m\phi-Et/\hbar]}\left[F_+(r)+iG_-(r)\right] \\
            e^{i[(m+s)\phi-Et/\hbar]}\left[F_-(r)-iG_+(r)\right] \\
           \end{array}
         \right),
\end{equation}
where we define
\begin{equation}\label{F} 
F_\kappa (r)\equiv\bar{C}_\kappa (E+d_f E_0+\kappa m_0 c^2)r^{\vert\Gamma_\kappa\vert}e^{-\frac{\Lambda r^2}{2}}L^{\vert\Gamma_\kappa\vert}_n(\Lambda r^2), \ \ (\kappa=\pm 1),
\end{equation}
and
\begin{equation}\label{G} 
G_\kappa (r)\equiv\hbar c\lambda\bar{C}_\kappa r^{\vert\Gamma_\kappa\vert}e^{-\frac{\Lambda r^2}{2}}\left[\left(-\frac{s\Gamma_\kappa-\kappa\vert\Gamma_\kappa\vert}{r}+(1-\kappa)\Lambda r\right)L^{\vert\Gamma_\kappa\vert}_n(\Lambda r^2)-2\kappa\Lambda r L^{\vert\Gamma_\kappa\vert+1}_{n-1}(\Lambda r^2)\right].
\end{equation}

It should be noted that our Dirac spinor simultaneously incorporates the positive and negative values of the quantum number $m$, which does not happen, for example, in Ref. \cite{Villalba}. However, the temporal part (temporal phase factor) is equal, given by $e^{-iEt/\hbar}$ (this is the minimal requirement for stationary states, whether in relativistic or nonrelativistic quantum mechanics). From a practical point of view, one of the advantages of we have a spinor with this characteristic is the possibility of calculating the physical observables faster and more direct than if we had two spinors, one for each value of $m$.


\section{Nonrelativistic limit}\label{sec5}

Here, let's investigate the nonrelativistic limit (regime) of our results, which is basically the low-energy limit (where most quantum phenomena occur), and it occurs when the speed of light tends to infinity ($c\to\infty$). To achieve this, we can use a standard prescription (``recipe'') widely used in the literature to obtain the nonrelativistic limit of relativistic wave equations for massive particles (among the various prescriptions, this is the ``simplest and easiest''). Thus, in such a prescription we need to consider that most of the total energy of the system is concentrated in the rest energy of the particle \cite{Greiner,Andrade1,Bermudez1,Villalba}, consequently, it implies that we can write the relativistic energy $E$ as: $E=m_0 c^2+\varepsilon$, in which $m_0 c^2$ must satisfy two conditions, given by: $m_0 c^2\gg \varepsilon$ and $m_0 c^2\gg U$, respectively. Therefore, using this prescription in Eq. \eqref{15}, we get the following Pauli-like NCQHO (nonrelativistic NCDO) with EDM in the
presence of an external electromagnetic field
\begin{equation}\label{L1}
H^{2D}_{Pauli-like}\varphi^{NC}=\left[\bar{H}^{2D}_{NCQHO-like}-\lambda\Omega(\hbar\sigma_3+sL_z)-d_f E_0\right]\varphi^{NC}=\varepsilon\varphi^{NC},
\end{equation}
or with the temporal dependence (general case), as
\begin{equation}\label{L2}
H^{2D}_{Pauli-like}\Psi^{NC}_{Pauli-like}=\left[\bar{H}^{2D}_{NCQHO-like}-\lambda\Omega(\hbar\sigma_3+sL_z)-d_f E_0\right]\Psi^{NC}_{Pauli-like}=i\hbar\frac{\partial\Psi^{NC}_{Pauli-like}}{\partial t},
\end{equation}
where 
\begin{equation}\label{QHO2}
\bar{H}^{2D}_{NCQHO-like}=\lambda^2\left[\frac{{\bf p}^2}{2m_0}+\frac{1}{2}\left(\frac{m_0\Omega^2}{\lambda^2}\right){\bf r}^2\right]=\frac{{\bf p}^2}{2M}+\frac{1}{2}M\bar{\Omega}^2{\bf r}^2, \ \ (\bar{\Omega}^2\equiv \lambda^2\Omega^2),
\end{equation}
with $H^{2D}_{Pauli-like}$ being the Hamiltonian of the Pauli-like NCQHO (is not technically the Hamiltonian of the noncommutative Pauli oscillator \cite{Heddar}), $\bar{H}^{2D}_{NCQHO-like}$ is the NCQHO-like Hamiltonian (different from \eqref{QHO}), $M=M^{NC}\equiv\frac{m_0}{\lambda^2}$ is a ``NC effective mass'', $\Psi^{NC}_{Pauli-like}=e^{-i\varepsilon t/\hbar}\varphi^{NC}$ is the NC spinorial wave function (two-component NC Pauli-like spinor), where $s=+1$ describes a particle with spin ``up'' and $s=-1$ describes a particle with spin ``down''. Besides, the second term in \eqref{L2} is a constant which shifts all energy levels (but does not affect the eigenfunctions), and the third term is the spin-orbit coupling term \cite{Moshinsky}. Summarizing, the nonrelativistic limit of the DO in (2+1)-dimensions (commutative or NC) results in the 2DQHO with a strong spin-orbit coupling term with all levels shifted by the factor $\lambda\Omega$ \cite{Andrade1}. Also, comparing Eq. \eqref{L2} with the literature, we verified that in the absence of the EDM ($d_f=0$), or of the electromagnetic field (${\bf B}={\bf E}=0$), and of the NCPS ($\theta=\eta=0$), we obtain the Pauli-like QHO with $s=\pm 1$ \cite{Andrade1} and $s=+1$ (``without spin'') \cite{Bermudez2}. Already for $\omega=\theta=\eta=E_0=0$ and $s=+1$ (without spin), and considering the lower component of the spinor ($\kappa=-1$) with $e^{im\phi}$ (consistent from a nonrelativistic point of view), we obtain the Schrödinger equation (SE) for a neutral particle with EDM in the presence of an external magnetic field, where now the magnetic field is is written as ${\bf B}=\frac{\rho_m}{2}r\hat{e}_r$, with $\rho_m$ being a magnetic charge density \cite{Ribeiro}. Explicitly, this SE it is basically the nonrelativistic limit of Eq. \eqref{17} for $\kappa=-1$ with $e^{im\phi}$.

However, the Hamiltonian $H^{2D}_{NCQHO-like}$ with $d_f=0$ is not the same Hamiltonian of the NCQHO found in the literature \cite{Giri} (particular case). In fact, this is because we introduced the NCPS in the linear DO and not directly in the quadratic DO. For example, if we had introduced the NCPS (from Ref. \cite{Giri}) only after obtaining the quadratic DO (i.e., introducing the NCPS into Eq. \eqref{14}), we would obtain the following quadratic NCDO 
\begin{equation}\label{L3}
H_{NCDO}^{quadratic}\varphi^{NC}=\left[H^{2D}_{NCQHO-like}-\bar{\omega}(\hbar\sigma_3+s(1+\Theta\bar{\Theta})L_z-s\Theta{\bf p}^2-s\bar{\Theta}{\bf r}^2)\right]\varphi^{NC}=\bar{E}\varphi^{NC},
\end{equation}
where 
\begin{equation}\label{QHO3}
H^{2D}_{NCQHO-like}=\frac{{\bf p}^2}{2M_{\Theta}}+\frac{1}{2}M_{\Theta}\Omega^2_{\Theta,\bar{\Theta}}{\bf r}^2-S_{\Theta,\bar{\Theta}}L_z, \ \ \bar{E}\equiv\left[\frac{(E+d_f E_0)^2-m_0^2 c^4}{2m_0 c^2}\right],
\end{equation}
being
\begin{equation}\label{MO}
M_{\Theta}\equiv\frac{1}{\left(\frac{1}{m_0}+m_0\bar{\omega}^2\Theta^2\right)}, \ \ \Omega_{\Theta,\bar{\Theta}}\equiv\sqrt{\left(\frac{1}{m_0}+m_0\bar{\omega}^2\Theta^2\right)\left(m_0\bar{\omega}^2+\frac{\bar{\Theta^2}}{m_0}\right)},
\end{equation}
and
\begin{equation}\label{SO}
S_{\Theta,\bar{\Theta}}\equiv\left(m_0\bar{\omega}^2\bar{\Theta}+\frac{\bar{\Theta}}{m_0}\right), \ \ \bar{\omega}\equiv(\omega-\sigma\tilde{\omega}).
\end{equation}

With respect to the nonrelativistic limit this equation, we have
\begin{equation}\label{L4}
H_{Pauli-like}^{2D}\varphi^{NC}=\left[H^{2D}_{NCQHO-like}-\bar{\omega}[\hbar\sigma_3+(1+\Theta\bar{\Theta})L_z-\Theta{\bf p}^2-\bar{\Theta}{\bf r}^2]-d_f E_0\right]\varphi^{NC}=\varepsilon\varphi^{NC},
\end{equation}
where taking $d_f=0$ ($\bar{\omega}\to\omega$) with $s=+1$, it implies that the Hamiltonian $H^{2D}_{NCQHO-like}$ now is exactly the same Hamiltonian (NCQHO) found in Ref \cite{Giri}. In that way, it shows that by introducing the NCPS into linear or quadratic DO, the result may not be the same.

Thus, using the standard prescription again, but now in \eqref{spectrum}, we obtain as a result the following nonrelativistic energy spectrum (nonrelativistic energy levels) for the Pauli-type NCQHO with EDM in the presence of an external electromagnetic field
\begin{equation}\label{spectrum2}
\varepsilon_{n,m,s}=U+\hbar N\left(1+s\frac{(\omega-\sigma\tilde{\omega})}{\omega_\theta}\right)(\omega+s\omega_\eta-\sigma\tilde{\omega}),
\end{equation}
where
\begin{equation}\label{potentialenergy2}
U\equiv-d_f E_0=-\sigma\vert d_f\vert E_0, \ \ (\sigma=\pm 1),
\end{equation}
and
\begin{equation}\label{N2}
N=N_{eff}\equiv\left[2n+1-\kappa+\Big|m+s\frac{1-\kappa}{2}\Big|-s\left(m+s\frac{1-\kappa}{2}\right)\right]\geq 0.
\end{equation}

Based on the spectrum \eqref{spectrum2}, we get Table \eqref{tab3}, where we have four possible configurations for the spectrum depending on the values of $s$ and $m$ (analogous to the relativistic case for the particle), in which $\tilde{n}=n+\frac{1-\kappa}{2}$, and we define for simplicity that $\bar{B}=2\hbar$.
\begin{table}[h]
\centering
\begin{small}
\caption{Nonrelativistic energy spectra for the degenerate states of the particle.} \label{tab3}
\begin{tabular}{ccc}
\hline
Configuration & Nonrelativistic energy spectrum & Degeneracy \\
\hline
1 &$\varepsilon_{n,+}=U+\tilde{n}\bar{B}\left(1+\frac{(\omega-\sigma\tilde{\omega})}{\omega_\theta}\right)(\omega+\omega_\eta-\sigma\tilde{\omega})$ & infinite \\
2 &$\varepsilon_{n,m<0,+}=U+\bar{B}\left(1+\frac{(\omega-\sigma\tilde{\omega})}{\omega_\theta}\right)(\omega+\omega_\eta-\sigma\tilde{\omega})(\tilde{n}+\vert m+\frac{1-\kappa}{2}\vert)$ & finite \\
3 &$\varepsilon_{n,m\geq 0,-}=U+\bar{B}\left(1-\frac{(\omega-\sigma\tilde{\omega})}{\omega_\theta}\right)(\omega-\omega_\eta-\sigma\tilde{\omega})(\tilde{n}+\vert m-\frac{1-\kappa}{2}\vert)$ & finite \\
4 &$\varepsilon_{n,-}=U+\tilde{n}\bar{B}\left(1-\frac{(\omega-\sigma\tilde{\omega})}{\omega_\theta}\right)(\omega-\omega_\eta-\sigma\tilde{\omega})$ & infinite \\
\hline
\end{tabular}
\end{small}
\end{table}

We note that the spectrum \eqref{spectrum2}, or the spectra from Table \eqref{tab3}, has some similarities and some differences with the relativistic case. For example, unlike the relativistic case:
\begin{itemize}
\item the spectrum \eqref{spectrum2} can admit negative or positive energy states (since we can have $N=0$). However, as we have seen in the previous section, $U$ positive (negative) means an EDM antiparallel (parallel) to the electric field;
\item depends linearly (a linear function) on the quantum numbers $n$ and $m$;
\item depends quadratically (a quadratic function) on the frequencies $\omega$ and $\tilde{\omega}$ (as we saw, both are ``variable interchangeable''); 
\item the parameter $\kappa$ describes a particle with spin ``up'' ($s=\kappa=+1$) or ``down'' ($s=\kappa=-1$).
\end{itemize}

Now, similar to the relativistic case:
\begin{itemize}
\item the spectrum \eqref{spectrum2} also depends linearly on $U$ (however, the concept of asymmetric spectrum does not apply here), and is greater (for $\sigma=-1$) in the presence of such energy (we will see this soon via graphs);
\item also has a finite or infinite degeneracy (depending on the values of $s$ and $m$);
\item also remains quantized even in the absence of the DO ($\omega=0$) and of the EDM ($d_f=0$), or   of the DO and of the NCPS ($\theta=\eta=1$);
\item $\theta$ and $\eta$ must also satisfy: $\theta>\frac{2s\hbar}{(\sigma\tilde{\omega}-\omega)}$ and $\eta>2s\hbar m_0(\sigma\tilde{\omega}-\omega)$;
\item the maximal spectrum is also for $s=+1$ and $m<0$ with $\sigma=-1$ and $\kappa=\pm 1$ (config. 2), while the minimal spectrum is for $s=-1$ and $m<0$ with $\sigma=\kappa=+1$ (config. 4);
\item also increases as a function of $n$, $m$, $\omega$, $\tilde{\omega}$, and $\omega_\eta$, and decreases as a function of $\omega_\theta$ (we will see this soon via graphs).
\end{itemize}

Furthermore, comparing the spectrum \eqref{spectrum2}, or the spectra from Table \eqref{tab3}, with some works of the literature, we verified that in the absence of the EDM ($d_f=0$), and of the NCPS ($\theta=\eta=0$), we obtain the usual spectrum of the Pauli-like QHO for $\kappa=+1$ \cite{Andrade1}. It is interesting to note that for $d_f=\theta=\eta=0$ with $\kappa=+1$ and $sm\geq 0$, we obtain also the spectrum of the Pauli-like QHO in three dimensions for $j=l+1/2$ $(E^{2D}_{n,ms\geq 0,\kappa=+1}=E^{3D}_{n,j=l+1/2})$ \cite{Strange}. However, the spectrum for $j=l-1/2$ (spin down) is larger for the three-dimensional case $(E^{2D}_{n,ms\geq 0,\kappa=+1}<E^{3D}_{n,j=l-1/2}$ and $E^{2D}_{n,ms<0,\kappa=+1}<E^{3D}_{n,j=l-1/2})$, as it should be. So, for $j=l+1/2$ and $sm\geq 0$ the Pauli-like QHO does not distinguish a two-dimensional universe from a three-dimensional (an ``anomaly'' in this case). In addition, taking the nonrelativistic limit of the spectrum \eqref{EDMspectrum} with $\kappa=-1$, $s=-1$, $m-1\to m$ (comes from the fact that now the angular part of the wave function is given by $e^{im\phi}$), $U=0$ (without electric field), and ${\bf B}\to \frac{\rho_m}{2}r\hat{e}_r$, we obtain the nonrelativistic spectrum of a ``free'' neutral particle with EDM in the presence of an external magnetic field \cite{Ribeiro}. From the above, we clearly see that our nonrelativistic spectrum generalizes several particular cases of the literature.

\begin{figure}[!h]
\centering
\includegraphics[width=10cm]{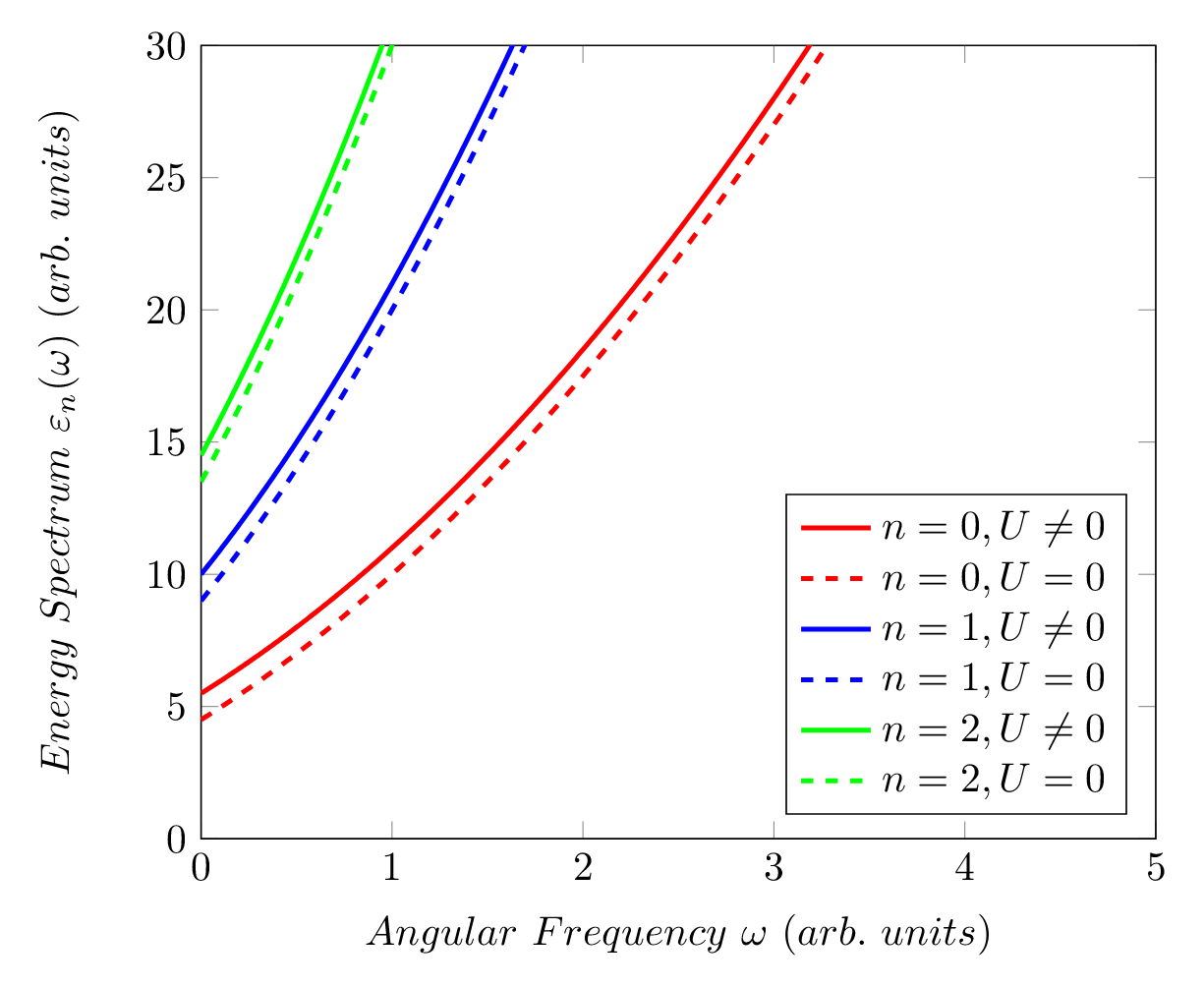}
\caption{Graph of $\varepsilon_n(\omega)$ versus $\omega$ for three different values of $n$ with $U\neq 0$ and $U=0$.}
\label{fig5}
\end{figure}

Now, let's analyze graphically the behavior of the nonrelativistic spectrum as a function of the four angular frequencies (for three actually, since $\omega\leftrightarrow\tilde{\omega}$ when $\sigma=-1$) for three different values of $n$ ($n=0,1,2$). Analogous to the relativistic case, here we will also choose the maximal spectrum, which is achieved for $s=+1$ and $m<0$
with $\sigma=-1$ and $\kappa=\pm 1$ (configuration 2 or 6). Therefore, we have Fig. \eqref{fig5}, where it shows the behavior of the energies as a function of $\omega$ for the ground state ($n=0$) and the first two excited states ($n=1,2$), with and without the presence of potential energy $U$ ($U\geq 0$), in which we take for simplicity that $\hbar=c=m_0=\vert d_f\vert=\Phi=\theta=\eta=1$.

\begin{figure}[!h]
\centering
\includegraphics[width=10cm]{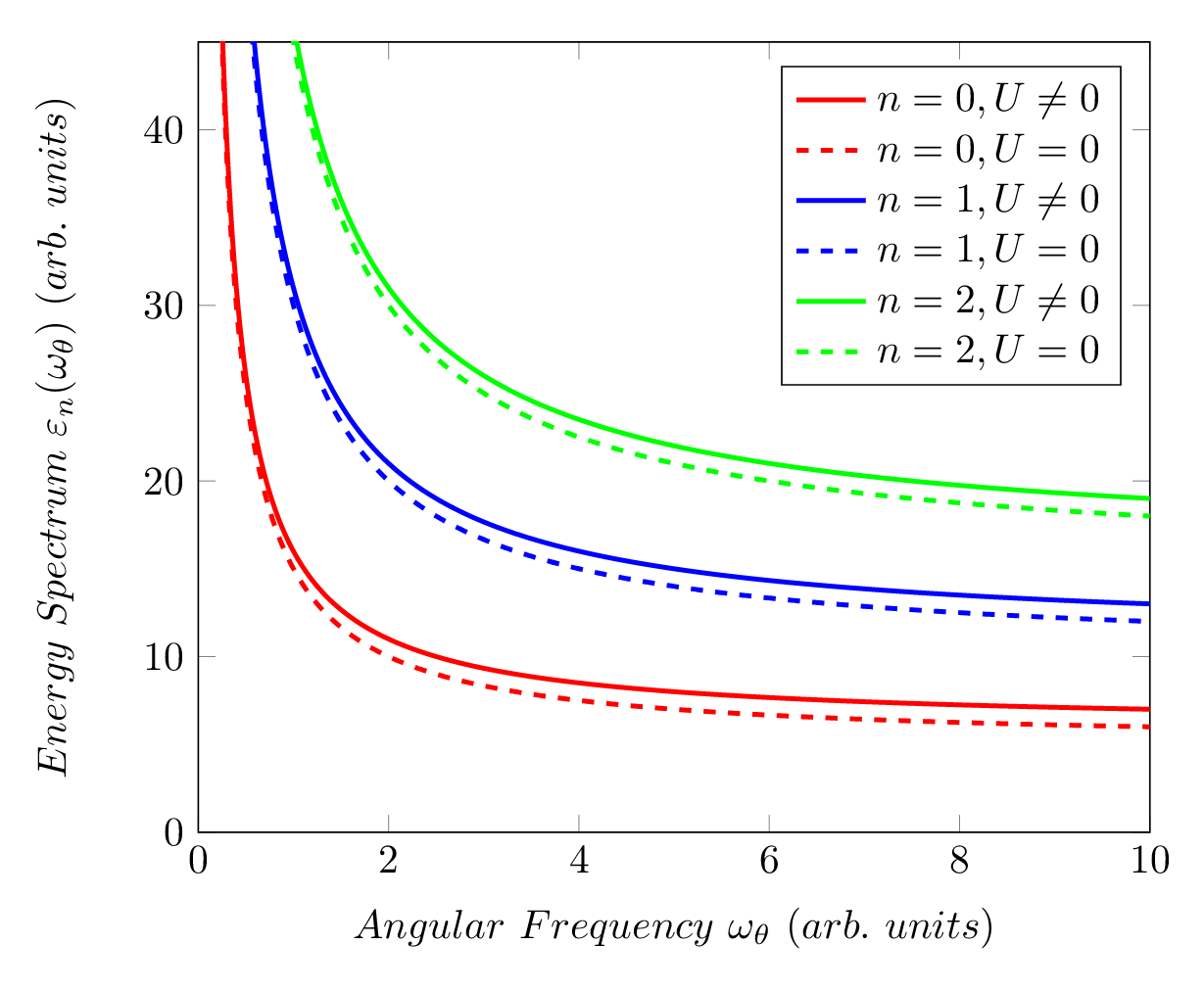}
\caption{Graph of $\varepsilon_n(\omega_\theta)$ versus $\omega_\theta$ for three different values of $n$ with $U\neq 0$ and $U=0$.}
\label{fig6}
\end{figure}

According to Fig. \eqref{fig5}, we see that the energies increase quadratically as a function
of $\omega$ (or $\tilde{\omega}$), and their values are larger with the increase of $n$ (as it should be) and in the presence of $U$. Then, analogous to the relativistic case, here the energy difference between two consecutive levels gets bigger ($\bigtriangleup E_n$ increases ``aggressively") as $\omega$ increases (either for the case $U\neq 0$ or $U=0$). In Fig. \eqref{fig6}, we have the behavior of the energies as a function of $\omega_\theta$ with and without the presence of $U$ ($U\geq 0$), where $n=0,1,2$, and $\hbar=c=m_0=\Phi=\omega=\eta=1$. According to this figure, we see that the energies decrease as a function of $\omega_\theta$, but increase as a function of $\theta$, where their values are larger with the increase of $n$ (as it should be) and in the presence of $U$. However, unlike the relativistic case, here this increase or decrease is more ``gentle'' (less ``aggressive''). Now, unlike the graph in Fig. \eqref{fig5}, here the energy levels have a nearly constant spacing as $\omega_\theta$ increases ($\bigtriangleup E\simeq const.$).

\begin{figure}[!h]
\centering
\includegraphics[width=10cm]{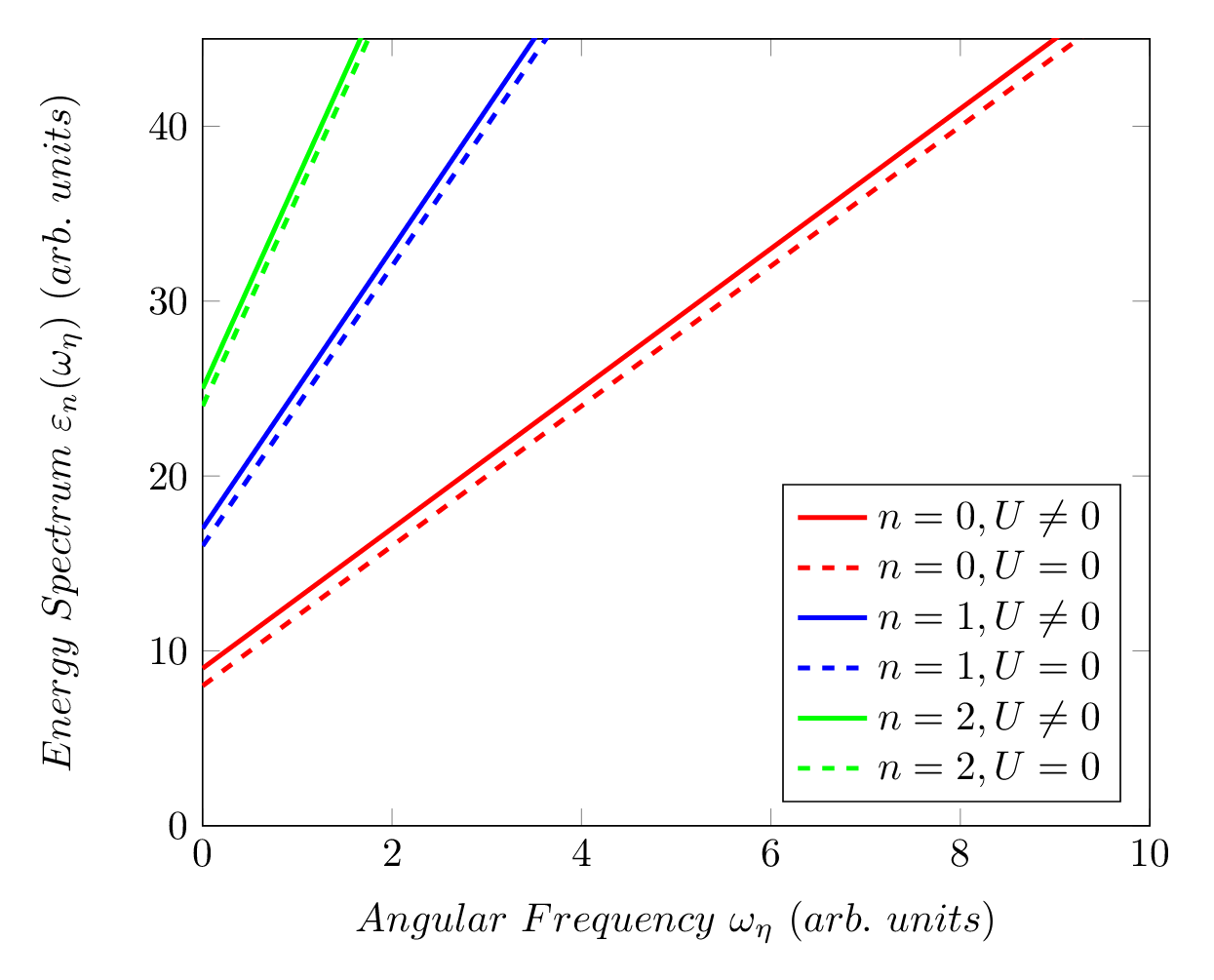}
\caption{Graph of $\varepsilon_n(\omega_\eta)$ versus $\omega_\eta$ for three different values of $n$ with $U\neq 0$ ($E_0=1$) and $U=0$ ($E_0=0$).}
\label{fig7}
\end{figure}

Already in Fig. \eqref{fig7}, we have the behavior of the energies as a function of $\omega_\eta$ with and without the presence of $U$ ($U\geq 0$), where $n=0,1,2$, and $\hbar=c=m_0=\Phi=\omega=\theta=1$. According to this figure, we see that the energies increase (linearly) as a function of $\omega_\eta$ (or $\eta$), where their values are larger with the increase of $n$ (as it should be) and in the presence of $U$. However, unlike the relativistic case, here this increase is more``aggressive''. Besides, analogous to the graph in Fig. \eqref{fig5}, here the energy levels also become more spaced out as $\omega_\eta$ increases ($\bigtriangleup E$ increases ``aggressively").

To end this section, let's now obtain the NC spinorial wave function (NC Pauli-like spinor) for the nonrelativistic bound states of the system. In particular, this function can be obtained in two different ways (but equivalent), namely: solving directly Eq. \eqref{L1}, or starting directly from the function \eqref{spinor3}. Here, we chose this second option because it is the least labor-intensive way. Therefore, we have the following NC spinorial wave function
\begin{equation}\label{spinor7}
\Psi^{NC}_{Pauli-like}=\left(
           \begin{array}{c}
            C'_+ e^{i[m\phi-\varepsilon t/\hbar]}r^{\vert m\vert}e^{-\frac{\Lambda r^2}{2}}L^{\vert m\vert}_n(\Lambda r^2) \\
            C'_- e^{i[(m+s)\phi-\varepsilon t/\hbar]}r^{\vert m+s\vert}e^{-\frac{\Lambda r^2}{2}}L^{\vert m+s\vert}_n(\Lambda r^2) \\
           \end{array}
         \right),
\end{equation}
where $C'_\kappa>0$ are new (nonrelativistic) normalization constants.

\section{Conclusion}\label{sec6}

In this paper, we investigate the bound-state solutions of the NCDO with a (permanent) EDM in the presence of an external electromagnetic field in the (2+1)-dimensional Minkowski spacetime. With respect to the electromagnetic field, we consider a radial magnetic field and linear at $r$ (or $x$ and $y$) generated by anti-Helmholtz coils, and the uniform electric field of the Stark effect (a uniform electric field in the z-direction). Defining the original Dirac spinor in terms of another spinor via a ``Dirac operator'' (has a similar form to the NCDO), we get at a second-order differential equation (radial or quadratic NCDO). Next, we introduce a new variable in this equation (via variable change), and then we analyze the asymptotic behavior of the resultant equation for $\rho\to 0$ and $\rho\to\infty$, we obtain as results a generalized Laguerre equation. From this equation, we determine the bound-state solutions of the system, given by the two-component NC Dirac spinor and the relativistic energy spectrum (relativistic energy levels). 

So, we verify that such a spinor is written in terms of the generalized Laguerre polynomials, and such a spectrum (negative and positive energies) is a linear function on the potential energy $U$ (generated by the interaction of the EDM with the electric field), and also depends explicitly
on the quantum numbers $n$ and $m$ (radial and magnetic quantum number), spin parameter $s$ (for a spin ``up'' or ``down''), and of four angular frequencies, given by: $\omega$, $\tilde{\omega}$, $\omega_\theta$, and $\omega_\eta$, respectively. In particular, $\omega$ is the well-known frequency of the NCDO, $\tilde{\omega}$ is a type of ``cyclotron frequency'', which is directly proportional to the EDM $d_f$ and the magnetic field $B_0$, and $\omega_\theta$ and $\omega_\eta$ are the NC frequencies of position and momentum, where $\omega_\theta$ is inversely proportional to the position NC parameter $\theta$, and $\omega_\eta$ is directly proportional to the momentum NC parameter $\eta$. In addition, some characteristics of the spectrum are: is a asymmetric spectrum due to the existence of $U$ (the energies of the particle and antiparticle are not equal), still remains quantized even in the absence of the DO ($\omega=0$) and of the EDM ($d_f=0$), or of the DO and of the NCPS ($\theta=\eta=1$), has a finite or infinite degeneracy (depending on the values of $s$ and $m$),  and the maximal values (highest allowed energies) are reached for $s=+1$ and $m<0$ with $d_f<0$ (regardless of the spinorial component chosen), while the minimal values (lowest allowed energies) are reached for $s=-1$ and $m<0$ with $d_f>0$ (for the upper spinorial component).

Furthermore, in order to obtain more physical information, we also analyze graphically the behavior of the spectrum as a function of the four angular frequencies for three different values of $n$. For example, in the graph of $\vert E^{\pm}(\omega)\vert$ versus $\omega$, we verify that the energies increase linearly as a function of $\omega$. On the other hand, the graph of $\vert E^{\pm}(\tilde{\omega})\vert$ versus $\tilde{\omega}$ (we omit this graph for simplicity) is identical to the of $\vert E^{\pm}(\omega)\vert$ versus $\omega$, and as $\tilde{\omega}\propto B_0$, implies that the energies increase linearly as a function of the magnetic field. In fact, the behavior between both graphs happens because these two frequencies are ``indistinguishable or interchangeable variables'' ($\omega\leftrightarrow\tilde{\omega}$), that is, we can exchange one frequency for the other and the graphs will be the same (for $d_f<0$). Consequently, from this we can give another interpretation for the origin of the nonminimal coupling of the DO, which would be through a neutral fermion with (negative) EDM interacting with a radial magnetic field and linear at $r$. Now, in the graph of $\vert E^{\pm}(\omega_\theta)\vert$ versus $\omega_\theta$ we verify that the energies decrease as a function of $\omega_\theta$, however, increase as a function of $\theta$. Already in the graph of $\vert E^{\pm}(\omega_\eta)\vert$ versus $\omega_\eta$, we verify that the energies increase as a function of $\omega_\eta$ (or $\eta$). Besides, in all these graphs we also verify that the energies of the particle are always greater than those of the antiparticle (but both increase as $n$ increases), where the energies of the particle (antiparticle) are larger (smaller) in the presence of $U$. Thus, comparing our relativistic spectrum with some other works, we verify that our results generalize several particular cases of the literature.

Finally, we also study the nonrelativistic limit (low-energy limit) of our results, where such a limit is obtained when consider that most of the total energy of the system is concentrated in the rest energy of the particle (it is a good standard prescription to obtain the nonrelativistic limit of relativistic wave equations). Therefore, as a direct consequence of this limit, we get the Pauli-like NCQHO in two-dimensions (2D), which is basically the NCQHO with a strong spin-orbit coupling term and a constant which shifts all energy levels, and whose solution is the NC spinorial wave function (NC Pauli-like spinor). Furthermore, an interesting fact is that in the absence of the NCDO ($\omega=\theta=\eta=0$), we obtain the 2DSE for a neutral particle with EDM in the presence of an external magnetic field. With respect to the nonrelativistic spectrum, we note that such a spectrum has some similarities and some differences with the relativistic case (for the particle). For example, unlike the relativistic case, the nonrelativistic spectrum can admit negative or positive energy states (there is no connection to an antiparticle), depends linearly of $n$ and $m$, and depends quadratically of $\omega$ (or $\tilde{\omega}$). Now, similar to the relativistic case, the nonrelativistic spectrum depends linearly on $U$, has a finite or infinite degeneracy, still remains quantized even in the absence of the DO and of the EDM (or of the DO and of the NCPS), and increases as a function of $n$, $m$, $\omega$, $\tilde{\omega}$, and $\omega_\eta$, and decreases as a function of $\omega_\theta$. So, comparing also our nonrelativistic spectrum with some other works, we verify that our results generalize several particular cases of the literature.

\section*{Acknowledgments}

The authors would like to thank the Funda\c c\~ao Cearense de Apoio ao Desenvolvimento
Cient\'ifico e Tecnol\'ogico (FUNCAP), the Coordena\c c\~ao de Aperfei\c coamento de Pessoal de N\'ivel Superior (CAPES), and the Conselho Nacional de Desenvolvimento Cient\'ifico e Tecnol\'ogico (CNPq) for financial support.


\end{document}